\documentclass[
 reprint,
 superscriptaddress,
 amsmath, amssymb,
 aps,
 pra,
 longbibliography
]{revtex4-2}

\usepackage{graphicx} 
\usepackage{dcolumn}   
\usepackage{bm}
\usepackage[margin=2.0cm]{geometry}
\usepackage{subfigure}
\usepackage{color}
\usepackage{natbib}
\usepackage{amsmath}
\usepackage{amsfonts}
\usepackage{amssymb}
\usepackage{braket}
\usepackage{bigints}
\usepackage{mathtools}
\usepackage[makeroom]{cancel}
\usepackage{dsfont}

\DeclareMathOperator{\Tr}{Tr}

\usepackage{xcolor}

\definecolor{darkgreen}{rgb}{0.0, 0.5, 0.0}

\usepackage[T1]{fontenc}

\usepackage{hyperref}
\hypersetup{
    colorlinks=true,
    linkcolor=blue,
    citecolor=blue,
    urlcolor=blue
}

\begin{document}

\preprint{APS/123-QED}

\title{Measurement compatibility in multiparameter quantum interferometry}

\author{Jayanth Jayakumar}
\affiliation{Faculty of Physics, University of Warsaw, ul.\ Pasteura 5, 02-093 Warsaw, Poland}
\affiliation{Faculty of Mathematics, Informatics and Mechanics, University of Warsaw, ul.\ Banacha 2, 02-097 Warsaw, Poland}
\author{Marco Barbieri}
\affiliation{Dipartimento di Scienze, Universit`a degli Studi Roma Tre, Via della Vasca Navale 84, 00146 Rome, Italy}
\author{Magdalena Stobińska}
\affiliation{Faculty of Mathematics, Informatics and Mechanics, University of Warsaw, ul.\ Banacha 2, 02-097 Warsaw, Poland}

\date{\today}

\begin{abstract}
The Cram\'er-Rao bound captures completely the performance of single-parameter quantum sensors. On the other hand, its extension to multiple parameters demands more caution. Different aspects need to be captured at once, including, critically, compatibility. In this article we consider  compatibility  in quantum interferometry for an important class of probe states, measured by double homodyne or photon counters, standard benchmarks for these applications. We include the presence of loss and phase diffusion in the estimation of a phase. Our results illustrate how different weighting of the precision on individual parameters affects their compatibility, adding to the list of considerations for quantum multiparameter estimation.
\end{abstract}

\maketitle

\section{\label{sec:level1}Introduction}

Quantum sensing holds promising perspectives for the observation of fragile materials with a reduced footprint~\cite{giovannetti2004quantum,paris2009quantum,demkowicz2015quantum,mauranyapin2022quantum}. Optimising the state of the probes and their measurement down to their quantum properties,  strategies capable of enhanced precision can be conceived~\cite{caves1981quantum,bollinger1996optimal,branford2019quantum,atkinson2021quantum,pezze2008mach,seshadreesan2011parity,d1995optimized}. The actual implementation may then be confronted with further constraints, such as the availability of components, or trade-offs in their performance. The presence of undesired effects is almost always unavoidable, and these need to be considered in any realistic sensor~\cite{banaszek2009quantum,maccone2011beauty,dorner2009optimal,demkowicz2009quantum,cable2010parameter,knysh2011scaling,escher2011general,kolodynski2014precision}. Embracing a multiparameter setting, in which target and nuisance parameters are estimated at once, has been proposed as a way of making the estimation more robust~\cite{nichols2018multiparameter,crowley2014tradeoff,altorio2015weak,vidrighin2014joint,szczykulska2017reaching,jayakumar2024quantum}. 

This advantage, however, comes at the cost of apportioning the available information content of the probes among the different parameters~\cite{lu2021incorporating,kull2020uncertainty}. In fact, the optimal measurements pertaining to individual parameters may not commute, preventing  joint optimal estimation. 
At the preparation level, one needs to consider whether a single probe state exists that is optimal for the estimation of all parameters~\cite{albarelli2022probe,hayashi2024finding,ohno2025simultaneous}.
Therefore, \emph{compatibility} becomes an outstanding aspect in the multiparameter scenario~\cite{ragy2016compatibility}. 

A possible way to capture this aspect demands comparing two information metrics: on the one side, the quantum Fisher information matrix~\cite{helstrom1969quantum,braunstein1994statistical,liu2020quantum}, on the other, Holevo information~\cite{holevo2011probabilistic,hayashi2008asymptotic}. The former quantifies the achievable variances assuming optimal measurement for each parameter. The latter, instead, develops on more sophisticated tools and delivers a bound on the variance that can be attained in principle, by a collective measurement on a large number of probes~\cite{roccia2017entangling,parniak2018beating,conlon2023approaching,hou2018deterministic,yuan2020direct,conlon2023discriminating}. The bound from the Fisher information, known as the quantum Cram\'er-Rao bound (QCRB) $C_Q$~\cite{helstrom1967minimum,helstrom1968minimum,yuen1973multiple}, then assumes more relaxed conditions which may never be met in the experiment, hence the Holevo Cram\'er-Rao bound (HCRB) $C_H$ is considered the most relevant scalar bound in this scenario~\cite{hayashi2005asymptotic}. Since $C_H$ is at most $2C_Q$~\cite{carollo2019quantumness,tsang2020quantum,demkowicz2020multi}, the quantity $C_H/C_Q$, introduced by Belliardo and Giovannetti, can be taken as an indicator quantifying the intrinsic compatibility of the problem~\cite{belliardo2021incompatibility}. 

The HCRB is established taking the most powerful possible measurements~\cite{guctua2006local,kahn2009local,massar1995optimal,yamagata2013quantum,yang2019attaining}, but there exist instances in which simpler strategies are sufficient to closely saturate it~\cite{roccia2017entangling,vidrighin2014joint}. However, these are rather exceptional cases, and the typical situation in the typical times is that feasible measurements, foremost non-collective ones, are far from its saturation~\cite{das2025holevo,conlon2208gap}, hence the importance of finding relevant figures of merit. In this article we draw inspiration from the Belliardo-Giovannetti ratio to devise a quantity that captures compatibility in a quantum interferometer. Here, we consider the double homodyne measurement and photon counting performed on an important class of two-mode multiphoton state with a fixed photon number as the probe~\cite{Holland93Interferometric}. Our results extend previous work~\cite{jayakumar2024quantum} devoted to the same scheme that demonstrated that these probes are not subject to an informational trade-off relation~\cite{massar1995optimal,vidrighin2014joint}  holding for two-dimensional probes. This article, instead, addresses the distinct issue of compatibility. The examples are physically motivated and represent instances of strong fundamental incompatibility (phase and loss estimation~\cite{crowley2014tradeoff}) and weak fundamental compatibility (phase and dephasing~\cite{vidrighin2014joint}). Our present findings illustrate how attributing different weights to the different parameters translates to their compatibility when phase is estimated together with loss or phase diffusion. Our results provide guidelines for inspecting the important issue of compatibility in practical cases.
    
\section{Background}

\subsection{The setup}\label{ssa}

Many optical sensors rely on the reading of a phase. These can be thus modelled as a Mach-Zehnder interferometer (MZI), in which the main aim is to estimate the relative phase between its two arms. In addition, for a comprehensive analysis, we should also include the presence of loss in both arms of the MZI, as well as some dephasing accounting for random fluctuations of the phase at faster time scales than the detection. These represent the main source of nuisance in realistic quantum interferometric sensors.  

The values of the noise parameters are often unknown at the time of measurement, and must be estimated along with the phase, since they are associated to the presence of the sample. Therefore, our measurements should be sensitive to noise parameters as well: this is particularly relevant for those ranges in which phase sensitivity is affected the most. This strategy is known to make the estimation more reliable~\cite{roccia2018multiparameter,belliardo2024optimizing}, however, the maximal sensitivity on each parameter is affected by their reciprocal interplay. This is a typical instance even in classical multiparameter estimation, made more ravelled by the intricacies of quantum measurements. These aspects are quantitatively captured by the HCRB, $C_H$, against which the actual sensitivity of a specific measurement scheme should be benchmarked. The key aspect is then understanding how its attainability is hampered by the actual performance of the measurement. This can be understood as the result of correlations of the parameters in the outcome probabilities, that are different from those predicated at the quantum limit.

We consider as the input, probe states in the form of the generalized Holland-Burnett (gHB) and the Holland-Burnett (HB) states, which are a class of experimentally relevant two-mode fixed photon number states~\cite{Holland93Interferometric,datta11imperfect,thekkadath20quantumenhanced,you2021scalable}. Given the distinct physical origin of the types of noise, we analyze the sensitivities of measurements considering the joint estimation of phase with each noise parameter separately. The gHB states are created by the action of a balanced beam splitter on a two-mode Fock state $\ket{n,N-n}$, namely 

\begin{align}
   |\Psi_{\mathrm{gHB}}(n,N-n)\rangle &{}= \mathcal{U}_{\mathrm{BS}}\ket{n,N-n} \\
   & =\sum_{p=0}^N \mathcal{A}_N(n,p)\ket{p, N-p}
   \label{eqn:ghbstate}
\end{align} 
where $\mathcal{U}_{\mathrm{BS}}=\exp{[-i\frac{\pi}{4}(a'^{\dagger}b' + b'^{\dagger} a')]}$ is the beam splitter unitary operator; $a',b'$ and $a'^{\dagger},b'^{\dagger}$ are the annihilation and creation operators, respectively, corresponding to the input modes of the beam splitter; and\\ $\mathcal{A}_{N}(n,p)=(-1)^n \sqrt{2^{-N} \binom{N}{n} \binom{N}{p}} \,
{}_2F_1\left(-n,-p;-N;2\right)$ are the Kravchuk coefficients~\cite{mycroft2023proposal}. The HB state is obtained as a special case of the gHB state at $n=N/2$ i.e., when we interfere equal numbers of photons on a balanced beam splitter.

In the MZI, the input gHB state goes through a phase-shift operation in which photons in one arm acquire a relative phase $\phi'$ with respect to the other. This is followed by an operation that introduces loss of photons in both arms modeled by beam splitters of transmissivities $\eta_a$ and $\eta_b$. Furthermore, the phase diffusion operation is introduced by allowing the values of the acquired phase to vary randomly according to a Gaussian probability distribution $p_{\phi,\Delta}(\phi')$ of mean $\phi$ and standard deviation $\Delta$. In other words, we take a weighted average of the lossy state over all phase values with respect to $p_{\phi,\Delta}(\phi')$. All of these operations can be condensed into the equation

\begin{align}\label{eqn:condensed}
    &\rho^{\mathrm{gHB}}_{\phi,\eta_a,\eta_b,\Delta}(n,N-n) 
     = \int_{-\infty}^{\infty} p_{\phi,\Delta}(\phi') \bigg( \sum_{k=0}^{N} \sum_{l=0}^{N-k} \\
    &\quad K_{b,l,\eta_b}\notag K_{a,k,\eta_a}\mathcal{U}_{\phi'}\rho^{\mathrm{gHB}}\mathcal{U}^{\dagger}_{\phi'} K^{\dagger}_{a,k,\eta_a} K^{\dagger}_{b,l,\eta_b} \bigg)\notag
    \, d\phi',
\end{align}
where $\mathcal{U}_{\phi'}=e^{-i\hat{a}^{\dagger}\hat{a}\phi'}$ is the phase-shift operation and $\hat{a}$ is one of the output modes of the beamsplitter. If $k$ and $l$ photons are lost from the modes $a$ and $b$, respectively, then $K_{a,k,\eta_a}=\sqrt{\frac{(1-\eta_a)^k}{k!}}\sqrt{\eta^{\hat{a}^{\dagger}\hat{a}}_a}\hat{a}^k$, $K_{b,l,\eta_b}=\sqrt{\frac{(1-\eta_b)^l}{l!}}\sqrt{\eta^{\hat{b}^{\dagger}\hat{b}}_b}\hat{b}^l$ are the Kraus operators that make up the photon loss channel, and $p_{\phi,\Delta}(\phi')=\frac{1}{\sqrt{2\pi\Delta^2}}e^{-(\phi'-\phi)^2/2\Delta^2}$

After evaluating the operations in Eq.\,\ref{eqn:condensed}, we obtain 

\begin{multline}\label{eqn:outputprobe}
\rho^{\mathrm{gHB}}_{\phi,\eta_a,\eta_b,\Delta}(n,N-n){}\\
{}=\begin{aligned}[t]\sum_{k=0}^{N}\sum_{l=0}^{N-k}\sum_{p,q=k}^{N-l}
&\mathcal{C}^{N}_{\phi,\eta_a,\eta_b,\Delta}(n,p,q){}\\&{}\times\ket{p-k, N-p-l}\bra{q-k, N-q-l},\end{aligned}
\end{multline}

where $\mathcal{C}^{N}_{\phi,\eta_a,\eta_b,\Delta}(n,p,q)\\
=\mathcal{A}_N(n,p)\mathcal{A}_N(n,q)e^{-i(p-q)\phi - \frac{\Delta^2}{2}(p-q)^2}\sqrt{B^p_{kl}B^q_{kl}}$,\\ $B^p_{kl}=\binom{p}{k}\binom{N-p}{l}\eta^{p-k}_a(1-\eta_a)^{k}\eta^{N-p-l}_b(1-\eta_b)^{l}$ quantifies the modification of the probability amplitudes due to losses.

Fixed-photon number states are generally associated with a measurement of the photon number at the two outputs of the MZI. This option typically shows good performance in quantum phase estimation, however it requires detectors in cryogenic environments, which may not be suitable for all applications. In this work, we also consider the use of a homodyne detector on each arm~\cite{vidrighin2014joint,zhong2021double}. While this explicitly demands a phase reference, this requirement has fewer technical constraints, and offers \emph{per se} higher efficiencies. 

\subsection{Multiparameter Cram\'er-Rao bounds}

In a general framework, we consider the estimation of $p$ parameters, represented as a vector $\vec{\theta}=(\theta_1,\ldots,\theta_p)^\top$ belonging to the parameter space $\Theta\subset\mathbb{R}^{p}$. The gHB state $\ket{\Psi_{\mathrm{gHB}}}$ has dimension $d$ and lives in the Hilbert space $\mathcal{H}\equiv\mathbb{C}^{d}$. The corresponding density matrix (input probe state) $\rho_{\mathrm{gHB}}$ belongs to the space of $d\times d$ Hermitian operators, $\mathcal{L}(\mathcal{H})\equiv\mathbb{C}^{d\times d}$. Information about the parameters can be extracted using a suitable measurement $M$ characterized by the positive operator-valued measure (POVM) $\{\Pi_x\}$. Thus, the action of $M$ with respect to the output probe state $\rho^{\mathrm{gHB}}_{\vec{\theta}}$ gives rise to a probability distribution  $\Tr(\rho^{\mathrm{gHB}}_{\vec{\theta}}M)=p_{\vec{\theta}}(x)$. Classically, it is desirable for $p_{\vec{\theta}}(x)$ to be sensitive to small changes in the parameter values, and, consequently, this aspect is key to establishing a metric for the performance of the measurement on the state. Considering $\nu$ separable copies of the probe, which introduces another classical advantage, the performance of the measurement can be related to the precision of estimating the parameters through the Cram\'er-Rao bound (CRB) reading
\begin{equation}\label{crb}
    \Tr(W\Sigma)\geq\frac{1}{\nu}\Tr(W F^{-1}_C)=C_C(M,\,\rho^{\mathrm{gHB}\otimes \nu}_{\vec{\theta}},W,\vec{\theta})
\end{equation}
where $\Tr(W\Sigma)$ is the average cost,\\ $\Sigma_{i,j}=\int p_{\vec{\theta}}(x) (\Tilde{\theta}_{i}-\theta_i)(\Tilde{\theta}_{j}-\theta_j)dx$ is the covariance matrix of unbiased estimators $\Tilde \theta_i$ with $\Sigma_{i,i}$ representing the estimation precision for each parameter. $F_C$ is the Fisher information matrix (FIM) whose elements read $F_{C\,i,j}=\bigintsss \frac{1}{p_{\vec{\theta}}(x)}\bigg(\frac{\partial p_{\vec{\theta}}(x)}{\partial\theta_i}\bigg)\bigg(\frac{\partial p_{\vec{\theta}}(x)}{\partial\theta_j}\bigg) dx$. In particular, $F_{C\,i,i}$ is the Fisher information (FI) that can be associated to each parameter when the others are known to arbitrary precision. Finally, $W$ is a non-negative definite weight matrix belonging to the set of $p \times p$ real matrices $\mathcal{S}^{\mathbb{R}}(\Theta)\subset \mathbb{R}^{p\times p}$, providing a penalty for the inaccurate estimation for each parameter. When $\theta_i$ are the parameters of interest, then $W$ is diagonal.

The quantum aspect of estimation theory is now introduced in order to obtain a measurement-independent quantum Cram\'er-Rao bound (QCRB) which depends only on the probe state:

\begin{equation}\label{qcrb}
    \Tr(W\Sigma)\geq\frac{1}{\nu}\Tr(W F^{-1}_Q)=C_Q(\rho^{\mathrm{gHB}\otimes \nu}_{\vec{\theta}},W,\vec{\theta})
\end{equation}

where $F_Q$ is the quantum Fisher information matrix (QFIM) whose elements read
$F_{Q\,i,j}=\mathrm{Re}\big(\Tr\big(\rho^{\mathrm{gHB}\otimes\nu}_{\vec{\theta}}L_{\theta_i}L_{\theta_j}\big)\big)=\frac{1}{2}\Tr\big(\rho^{\mathrm{gHB}\otimes\nu}_{\vec{\theta}}\{L_{\theta_i},L_{\theta_j}\}\big)$, $L_{\theta_i}$ is the symmetric logarithmic derivative (SLD) for the parameter $\theta_i$. In particular $F_{Q\,i,i}$ is the quantum Fisher information (QFI) for each parameter. The QFI represents the amount of information about a parameter contained in the probe state. However, the QCRB does not provide a recipe for the optimal measurement, nor does it ensure it actually exists. The reason is that the optimal measurements associated with distinct parameters may not commute; hence, their joint estimation at the ultimate quantum precision may not be attained.

A more nuanced optimization of the covariance matrix gives rise to a bound that is informative and tighter than the QCRB known as the HCRB as follows.

\begin{equation}\label{hcrb}
    \Tr(W\Sigma)\geq C_H(\rho^{\mathrm{gHB}\otimes \nu}_{\vec{\theta}},W,\vec{\theta})
\end{equation}

The bound $C_H$ is obtained as the outcome of the following convex optimization problem:

\begin{equation}\label{eqn:hcrb1}
\begin{aligned}
    C_H(\rho^{\mathrm{gHB}\otimes \nu}_{\vec{\theta}},W,\vec{\theta}) &= \frac{1}{\nu} \min\limits_{\vec{X},V} \Big[ \Tr(WV) \\
    & \quad \begin{cases}
             V\geq Z(\vec{X})\\
             \Tr(\nabla\rho^{\mathrm{gHB}\otimes\nu}_{\vec{\theta}}\vec{X}^{\mathrm{T}})=\mathds{1}
    \end{cases}\Bigg]
\end{aligned}
\end{equation}

where $\vec{X}=(X_1,\ldots,X_p)^\top$ is a vector of $p$ Hermitian matrices such that $ X_i \in \mathcal{L}(\mathcal{H})$, $Z(\vec{X})=\Tr(\rho^{\mathrm{gHB}\otimes\nu}_{\vec{\theta}}\vec{X}\vec{X}^{\mathrm{T}})$ belongs to the set of $p \times p$ complex matrices $\mathcal{S}^{\mathbb{C}}(\Theta)\subset \mathbb{C}^{p\times p}$, $V$ is a $p \times p$ real matrix belonging to $\mathcal{S}^{\mathbb{R}}(\Theta)$, and $\Tr(\nabla\rho^{\mathrm{gHB}\otimes\nu}_{\vec{\theta}}\vec{X}^{\mathrm{T}})=\mathds{1}$ is the local unbiasedness condition. The HCRB captures information about fundamental incompatibility and is known to be attainable under very general conditions, which, however, may have limited relevance in the practical case. More details are presented in Appendix~\ref{app:hcrb}.

\subsection{Figures of merit for measurement compatibility}\label{sec:foms}

The two quantum bounds describe ultimate limits to the precision, with the HCRB being more pertinent in the practical case. This has led Belliardo and Giovannetti to introduce a figure of merit with the aim of quantifying compatibility as~\cite{belliardo2021incompatibility}
\begin{equation}\label{fom2}
    r_{BG} \big(\rho^{\mathrm{gHB}}_{\vec{\theta}},W,\vec{\theta}\big)=\frac{C_H\big(\rho^{\mathrm{gHB}}_{\vec{\theta}},W\big)}{C_Q\big(\rho^{\mathrm{gHB}}_{\vec{\theta}},W\big)}.
\end{equation}
This describes how close the best possible measurement scheme can get to the predicament of the QCRB, setting the {\it intrinsic} level of incompatibility in this estimation problem. In the analysis of practical cases, one is actually interested in understanding what extra cost is associated to the use of a specific measurement strategy, therefore a suitable figure is given by   
\begin{equation}\label{fom1}
    r^{C}_H\big(M,\rho^{\mathrm{gHB}}_{\vec{\theta}},W,\vec{\theta}\big)=\frac{C_C(M,\rho^{\mathrm{gHB}}_{\vec{\theta}},W)}{C_H\big(\rho^{\mathrm{gHB}}_{\vec{\theta}},W\big)}
\end{equation}
This quantifies the compatibility offered by a chosen measurement $M$. Note that the additivity property of $C_C$, $C_Q$, and $C_H$ renders these measures independent of $\nu$ when separable probe states are considered, and for our analysis, we deal with a single copy of the probe i.e., $\nu=1$. Hence, we have denoted $\rho^{\mathrm{gHB}\otimes\nu}_{\vec{\theta}}$ as $\rho^{\mathrm{gHB}}_{\vec{\theta}}$ in Eqs.\,\ref{fom2} and \ref{fom1}. Since $C_C\geq C_H \geq C_Q$, one can see that $r^C_H\in [1,\infty)$, with $r^C_H=1$ implying full measurement compatibility, where the measurement saturates the HCRB, while $r^C_H>>1$ indicates low compatibility.

In practice, one is often restricted to separable measurements, for which the HCRB may provide an optimistic bound to the error. With such a scheme, a more relevant benchmark is provided by the Nagaoka-Hayashi-Cram\'er-Rao bound (NHCRB), $C_{N}$~\cite{hayashi2005asymptotic,nagaoka2005new}, which  is saturated by the optimal separable measurement (see Appendix~\ref{app:nhcrb} for more details). For the purpose of assessing the performance of $M$, we thus define another figure
\begin{equation}\label{fom3}
    r^{C}_N\big(M,\rho^{\mathrm{gHB}}_{\vec{\theta}},W,\vec{\theta}\big)=\frac{C_C(M,\rho^{\mathrm{gHB}}_{\vec{\theta}},W)}{C_N\big(\rho^{\mathrm{gHB}}_{\vec{\theta}},W\big)}
\end{equation}

When the measurement $M$ is optimal, {\it i.e.} $C_C=C_N$, $r^C_H$ corresponds to
$r^{N}_H=C_N/C_H$, quantifying the advantage offered by the use of collective measurements for a given probe in the estimation problem. In particular, for the estimation of $p$ parameters, we have: $r^{N}_H\in[1,p]$ \cite{das2025holevo}.

The extent of measurement compatibility is also determined by the relative weights associated to the parameters, thus the scope of prioritising and penalising the error on one parameter over the other is relevant. Towards this goal, we work with the weight matrix $W(y)=\begin{bmatrix}
2 y & 0\\
0 & 2(1-y)
\end{bmatrix}$, $y\in[0,1]$. By this choice, we can interpolate from the identity matrix, corresponding to $y=0.5$, for an unbiased penalization, to the extremal cases $y=0$ and $y=1$, yielding rank-1 matrices that penalize the error of only one of the two parameters. This also allows us to obtain bounds on linear combinations of estimation errors of individual parameters.

As for the evaluation of $r^{C}_{H}$, an analytical computation of $C_C$ is not feasible due to the non-trivial structure of the double homodyne POVM (see Appendix~\ref{app:dhd}) and the noisy character of the output state (Eq.~\ref{eqn:outputprobe}). This is then evaluated numerically in a straightforward manner. $C_H$ is obtained as the solution of a convex optimization problem, allowing to cast Eq.~\ref{eqn:hcrb1} as a semidefinite program (SDP), that is efficiently computable~\cite{albarelli2019evaluating}. A closed-form solution, $\overline{C}_H$, only exists for the so-called D-invariant models~\cite{albarelli2020perspective}, to which our problem does not belong, with $C_H \leq \overline{C}_H$~\cite{tsang2020quantum,albarelli2020perspective} (see Appendix~\ref{app:hcrb}).
For these reasons, we resort to a numerical evaluation. We make use of the source code for the \texttt{HCRB} function from the \texttt{QuanEstimation} package \cite{zhang2022quanestimation}, which feeds the numerical model to the \texttt{CVXPY} modeling framework in Python. For gHB states, particularly those of higher dimensions, we have further improved and optimized this code and solved the SDP using the \texttt{MOSEK} solver \cite{mosek}. The optimized code is available at~\cite{Jayakumar_HCRB_for_gHB_2025}. The bounds $C_Q$ and $C_N$, used in $r_{BG}$ and $r^C_N$, respectively, are computed numerically using the \texttt{QuanEstimation} package.

\section{Results}
\subsection{Joint-estimation of phase and loss}\label{sec:2c1}

\begin{figure*}
\centering
\includegraphics[scale=0.5]{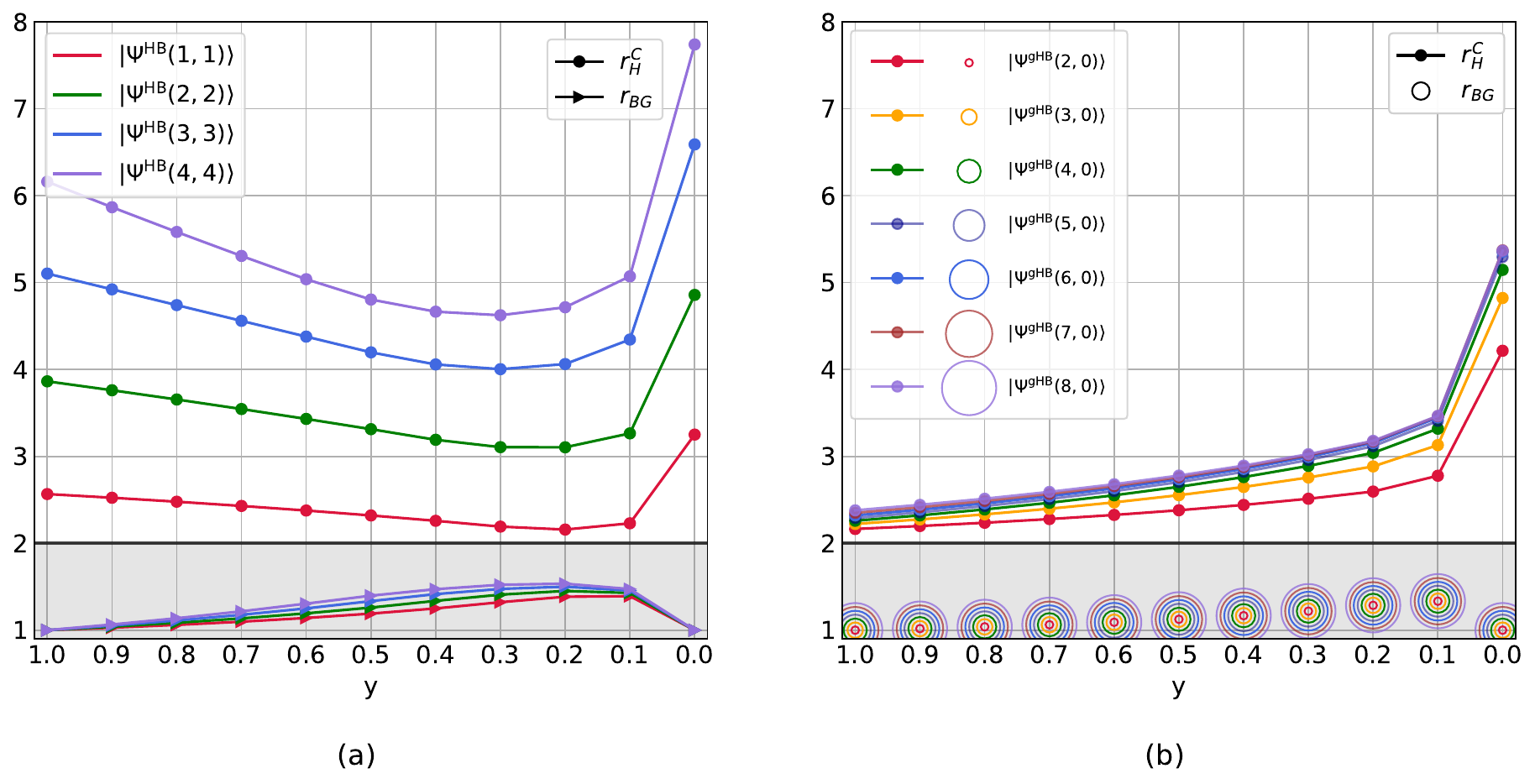}
\caption{\small{Plot of measurement compatibility measure $r^C_H$ and fundamental compatibility measure $r_{BG}$ versus the weights $y$ for the joint-estimation of phase and losses in one arm of the MZI. HB states with $N=2,4,6,8$ (panel (a)) and gHB states with $N=2,3,4,5,6,7,8$ (panel (b)), and double homodyne measurement are considered. In panel (b) the plotting of $r_{BG}$ employs concentric circles of varying colors to represent the near-perfect overlap of compatibilities for different values of $N$, as indicated by their common center. Note that in the following plots, such overlapping points are depicted in this manner. The curves are plotted at $\eta_a=0.5$ and $\Delta=0$. For HB states, the measurement compatibility decreases significantly as $N$ increases, while the fundamental compatibility decreases only slightly as $N$ increases.}}
\label{fig:dhd_oa}
\end{figure*}

The first application of the general setting described in Section 
~\ref{ssa} concerns the joint estimation of phase and loss in one arm of the interferometer~\cite{crowley2014tradeoff}. The corresponding output probe state $\rho^{\mathrm{gHB}}_{\phi,\eta_a}$ is obtained from Eq.\,\ref{eqn:outputprobe} by setting $\eta_b=1$ (no loss on the reference arm), $\Delta=0$ (no dephasing). Crucially, this output has a direct sum structure $\rho^{\mathrm{gHB}}_{\phi,\eta_a}(n,N-n)=\bigoplus_{k=0}^{N}p_{k}\ket{\xi_{\phi,\eta_a}(k)}\bra{\xi_{\phi,\eta_a}(k)}$, where each term $\ket{\xi_{\phi,\eta_a}(k)}=\newline\frac{1}{\sqrt{p_k}}\sum_{p=k}^{N}\mathcal{A}_N(n,p)e^{-ip\phi}\sqrt{B^p_k}\ket{p-k,N-p}$ is a pure state belonging to the subspace corresponding to the loss of $k$  photons, with a normalization factor or the associated probability, $p_k$ \cite{demkowicz2009quantum,crowley2014tradeoff}. These states also satisfy orthogonality $\braket{\xi_{\phi,\eta_a}(k),\xi_{\phi,\eta_a}(k')}=\delta_{kk'}$. Although we have computed the QCRB numerically here, the linearity of QFI on direct sums~\cite{helstrom1969quantum,fujiwara2001quantum} allows us, in principle, to calculate the QCRB analytically.

Fig.\,\ref{fig:dhd_oa} depicts the measurement compatibility $r^C_H$ and, for comparison, the fundamental compatibility measure $r_{BG}$ as a function of the weight  $y$ for our choice of double homodyne measurement on different probe states. The bounds depend on the value of the estimated parameter $\eta_a$, but are independent of $\phi$. Specifically, in panel (a), we consider HB states for $N=2,4,6,8$, in panel (b), we consider the gHB states: $\ket{\Psi_{\mathrm{gHB}}(N,0)}$ for $N=2,3,4,5,6,7,8$. We compute the CRB, QCRB, and HCRB at $\eta_a=0.5$ so that the compatibilities are studied in scenario of moderate loss. We remark that for this lossy model there is no advantage in the adoption of collective measurements, or, in different words, the parameter $r^N_H=C_N/C_H$ is 1 for all states and weights we consider. This is connected to the fact that the two parameters, phase and loss, are strongly incompatible: $\Tr(\rho^{gHB}_{\phi,{\eta_a}}[L_\phi,L_{\eta_a}])\neq 0$~\cite{albarelli2020perspective}. Notice that the extreme points $y=0$ and $y=1$ correspond to a single-parameter estimation, thus the notion of compatibility does not apply.

In panel \ref{fig:dhd_oa}(a), for all values of $N$,  $r^C_H$ shows a similar behaviour, with a minimum value achieved around $y=0.2$, corresponding to the maximal compatibility allowed by our choice of measurement. As $N$ increases, the overall magnitude of $r^C_H$ also increases, signalling that double homodyne becomes more inefficient at extracting information. We can thus investigate whether this is at least partly mirrored in the behaviour of the fundamental compatibility $r_{BG}$. This reaches the value 1 at the extremes, as expected for single-parameter problems, and exhibits a maximum around $y=0.2$, only weakly depending on $N$. On the other hand, fundamental incompatibility too increases with the number of photons, as $r_H^{C}$ does, although on a different scale.

In panel \ref{fig:dhd_oa}(b), there occurs an increase of $r^C_H$ as a function of $N$, but this dependency is much less pronounced. In contrast to panel (a), the best compatibility conditions for $r^C_H$ occur in proximity of pure phase estimation, close to $y=0.9$, corresponding to maximal measurement compatibility.  Interestingly, we find that, at the level of individual photon number, the behavior of $r^C_H$ mirrors that of $r_{BG}$. The latter also follows the behavior observed in panel (a). However, at the collective level, in contrast to measurement compatibility, the fundamental compatibility remains constant with the number of photons.

\begin{figure*}
\centering
\includegraphics[scale=0.5]{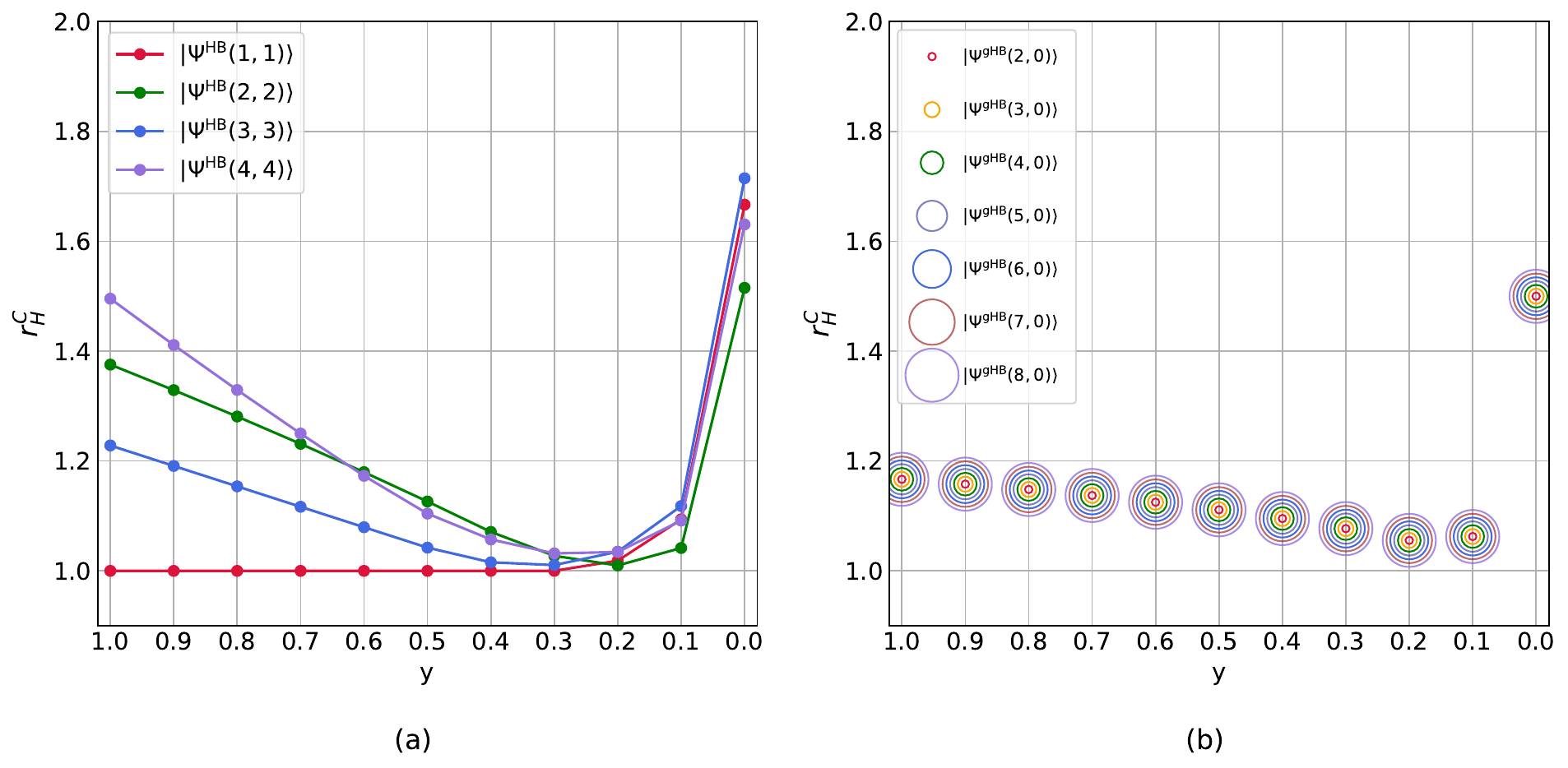}
\caption{\small{Plot of measurement compatibility measure $r^C_H$ versus the weights $y$ for photon counting for the joint-estimation of phase and loss in one arm of the MZI. HB states with $N=2,4,6,8$ (panel(a)) and gHB states with $N=2,3,4,5,6,7,8$ (panel(b)) are considered. Note that the measurement compatibility remains relatively high compared to that of the double homodyne measurement, with the highest compatibility observed at $N = 2$. The other known parameters are taken at the same values as those used in the assessment of the double homodyne measurement.}}
\label{fig:pn_oa}
\end{figure*}

Fig.~\ref{fig:pn_oa} reports the same analysis now carried out for photon counting performed at the two outputs of the MZI. Panel \ref{fig:pn_oa}(a) shows HB states with $N=2,4,6,8$ and panel \ref{fig:pn_oa}(b) shows gHB states with $N=2,3,4,5,6,7,8$. For HB states, we find that for $N=2$, $r^C_H\approx1$ up to $y=0.2$, thus showing good compatibility. On the other hand, for the remaining values of $N$ we considered, the trend becomes non-monotonic, with a minimum value of $r^C_H$ achieved near $y=0.3$. For gHB states, instead, the behaviour is independent on the photon number $N$, with a minimum found in the same region as in the previous case. For all those cases, the compatibility remains satisfactory when $y\in[0.9,0.1]$, as it stays below 1.5. This observation is in line with previous studies that have highlighted a better metrological performance of photon counting, a non-Gaussian measurement, on non-Gaussian probes~\cite{zhong2021double}.

\begin{figure*}
\centering
\includegraphics[scale=0.5]{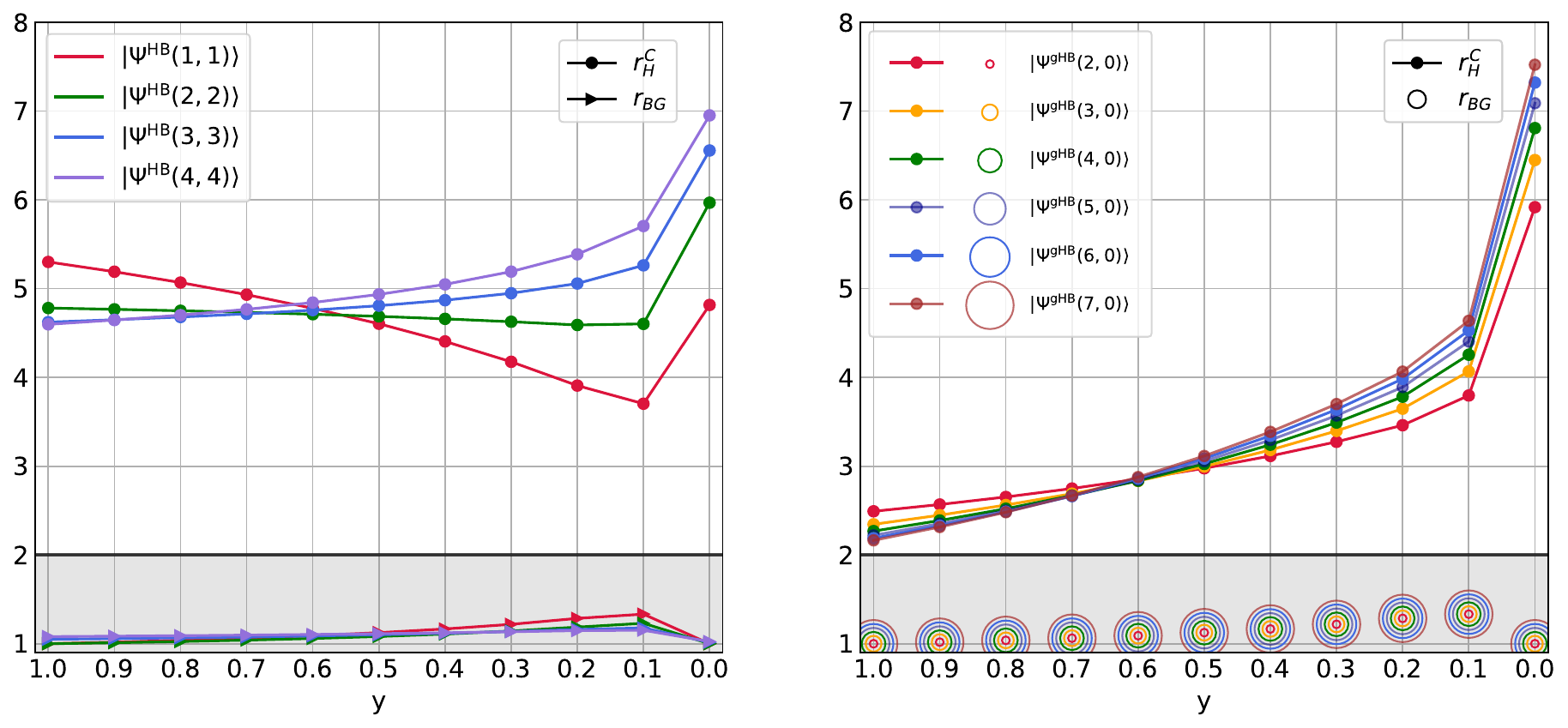}
\caption{\small{Plot of measurement compatibility measure $r^C_H$ and fundamental compatibility measure $r_{BG}$ versus the weights $y$ for the joint-estimation of phase and losses in one arm of the MZI with a known amount of loss on the reference arm. HB states with $N=2,4,6,8$ (panel (a)) and gHB states with $N=2,3,4,5,6,7$ (panel (b)), and double homodyne measurement are considered. The curves are plotted at $\eta_a=0.5$, $\eta_b=0.5$, and $\Delta=0$. Note that for HB and gHB states, the measurement compatibility exhibits an intersection at $y=0.6$, causing its behavior with respect to $N$ to change on either side of this point. In contrast, the fundamental compatibility increases slightly as $N$ increases.}}
\label{fig:dhd_ba}
\end{figure*}

\begin{figure*}
\centering
\includegraphics[scale=0.5]{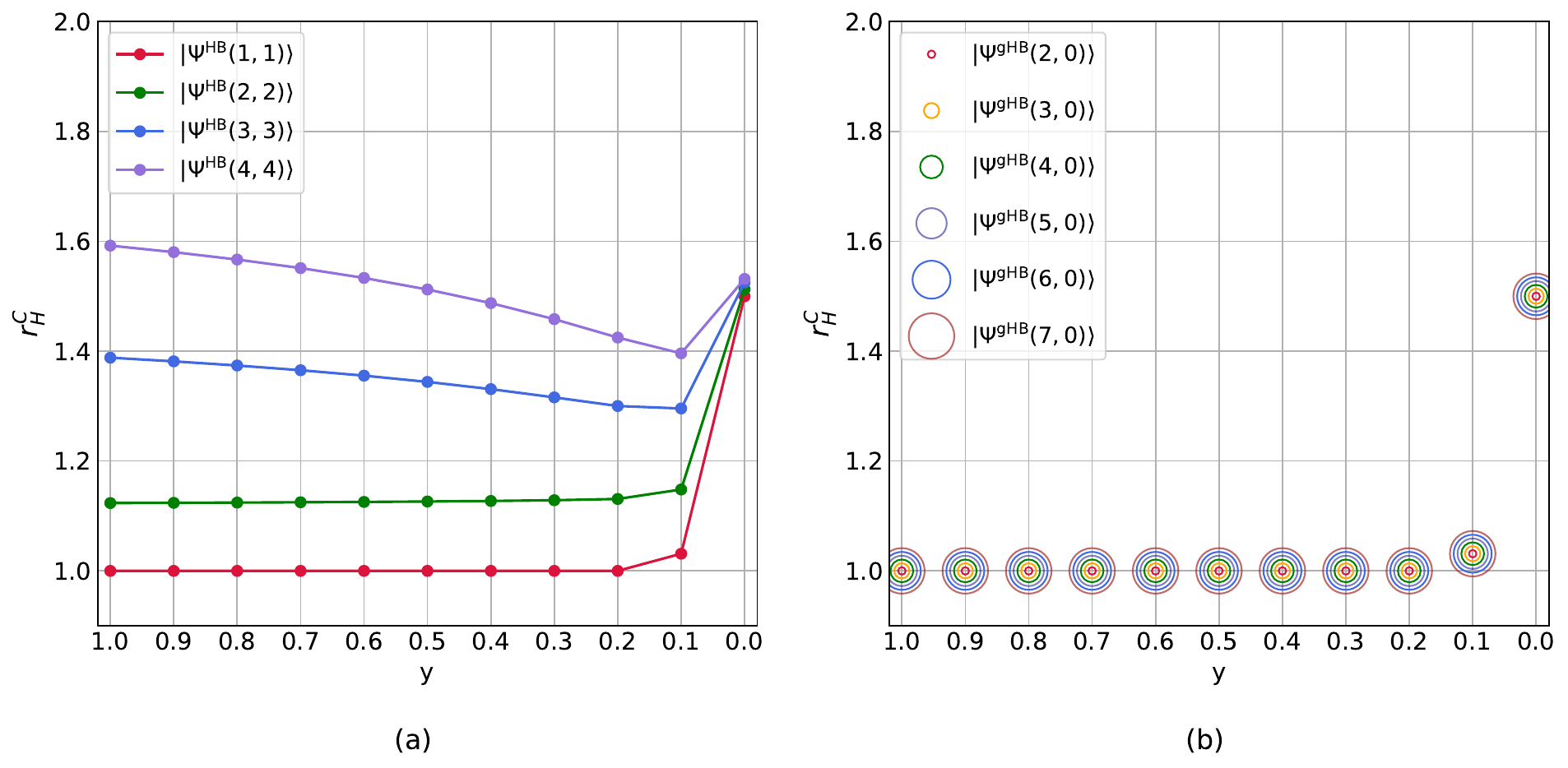}
\caption{\small{Plot of measurement compatibility measure $r^C_H$ versus the weights $y$ for photon counting for the joint-estimation of phase and losses in one arm of the MZI with a known amount of loss on the reference arm. HB states with $N=2,4,6,8$ (panel(a)) and gHB states with $N=2,3,4,5,6,7$ (panel(b)) are considered. Notably, the measurement compatibility remains high in this case as well, exceeding that of the double homodyne measurement, with optimal compatibility occurring at $N=2$. However, the compatibility strongly decreases as $N$ increases. The other known parameters are taken at the same values as those used in the assessment of the double homodyne measurement.}}
\label{fig:pn_ba}
\end{figure*}

We now extend our study to the joint estimation of phase and loss in one arm of the interferometer, while assuming a known amount of loss in the other arm. The corresponding output probe state $\tilde{\rho}^{\mathrm{gHB}}_{\phi,\eta_a}$ is obtained from Eq.\,\ref{eqn:outputprobe} by setting $\eta_b=0.5$ (50\% loss on the reference arm) and $\Delta=0$. The output can be expressed as $\rho^{\mathrm{gHB}}_{\phi,\eta_a}(n,N-n)=\newline\sum_{k=0}^{N}\sum_{l=0}^{N-k}p_{kl}\ket{\xi_{\phi,\eta_a}(k,l)}\bra{\xi_{\phi,\eta_a}(k,l)}$, where each term $\ket{\xi_{\phi,\eta_a}(k,l)}=\newline\frac{1}{\sqrt{p_{kl}}}\sum_{p=k}^{N-l}\mathcal{A}_N(n,p)e^{-ip\phi}\sqrt{B^{p}_{kl}}\ket{p-k,N-p-l}$ is a pure state conditioned on the number of lost photons, $k$ and $l$, in each arm \cite{demkowicz2009quantum}. In general, $\tilde{\rho}^{\mathrm{gHB}}_{\phi,\eta_a}$ does not have a direct sum structure since the states corresponding to the same total number of lost photons $m=k+l$ do not satisfy orthogonality. However, we note that, due to the convexity of QFI~\cite{helstrom1969quantum,fujiwara2001quantum}, one can still obtain an analytical upper bound to the QFI but it is beyond the scope of this article. We remark that in this case too the computation of the QFI and that of the QCRB could be performed analytically. However, the measurement compatibility requires the HCRB, and determining its analytical expression is challenging for reasons mentioned in Section~\ref{sec:foms}.

In Fig.\,\ref{fig:dhd_ba}, we plot $r^C_{H}$ and $r_{BG}$ as a function of $y$ for the double measurement on different probe states. Specifically, in panel (a), we consider HB states for $N=2,4,6,8$, in panel (b), we consider the gHB states: $\ket{\Psi_{\mathrm{gHB}}(N,0)}$ for $N=2,3,4,5,6,7$, and compute the CRB, QCRB, and HCRB at $\eta_a=0.5$. In panel \ref{fig:dhd_ba}(a), we observe different orderings in compatibility for phase and loss depending on the photon number $N$. When phase has more weight, $y\simeq 1$, increasing the photon number appears advantageous for $r^C_H$: one can attribute this to more distinct fringes in the probabilities of the two-homodyne outcomes. On the other hand, when loss has more weight, $y\simeq0$, lower $N$ results in better compatibility. As a consequence, the curves show an intersection around $y=0.6$. The inspection of the fundamental incompatibility via $r_{BG}$ reveals that this feature is specific to our choice of measurement, with higher $N$ allowing for better compatibility, at a difference with respect to the previous case (panel Fig.~\ref{fig:dhd_oa}(a)).  In panel \ref{fig:dhd_ba}(b), the same trends are found for the double homodyne measurement, with an intersection occurring around $y=0.6$.  Both the individual and the collective behaviours of $r_{BG}$ remain the same as in Fig.~\ref{fig:dhd_oa}(b). An investigation for generic loss reveals that the crossing point depends on the level of loss. Looking at the extreme cases, it is found that, for high losses on the reference arm ($\eta_b=0.1$),  $r^C_H$ decreases as $N$ increases regardless $y$, whereas for low losses ($\eta_b=1$), it increases with $N$ for all values of $y$ (see Appendix\,\ref{app:refarm}).

In Fig.~\ref{fig:pn_ba}, we investigate the measurement compatibility for photon counting for this setting. In particular, panel \ref{fig:pn_ba}(a) corresponds to HB states with $N=2,4,6,8$ and panel \ref{fig:pn_ba}(b) corresponds to gHB states with $N=2,3,4,5,6,7$. In panel \ref{fig:pn_ba}(a), we find that collectively, $r^C_H$ exhibits the same features with respect to $N$. For a given $N$, the minimum value is achieved at $y=0.9$ i.e., when the errors on loss bear little importance. However, with the increase in the number of photons, $r^C_H$ also increases. In panel \ref{fig:pn_ba}(b), $r^C_H$ remains constant and close to 1 for all values of $N$, except in proximity of $y=1$.

\begin{figure*}
\centering
\includegraphics[scale=0.55]{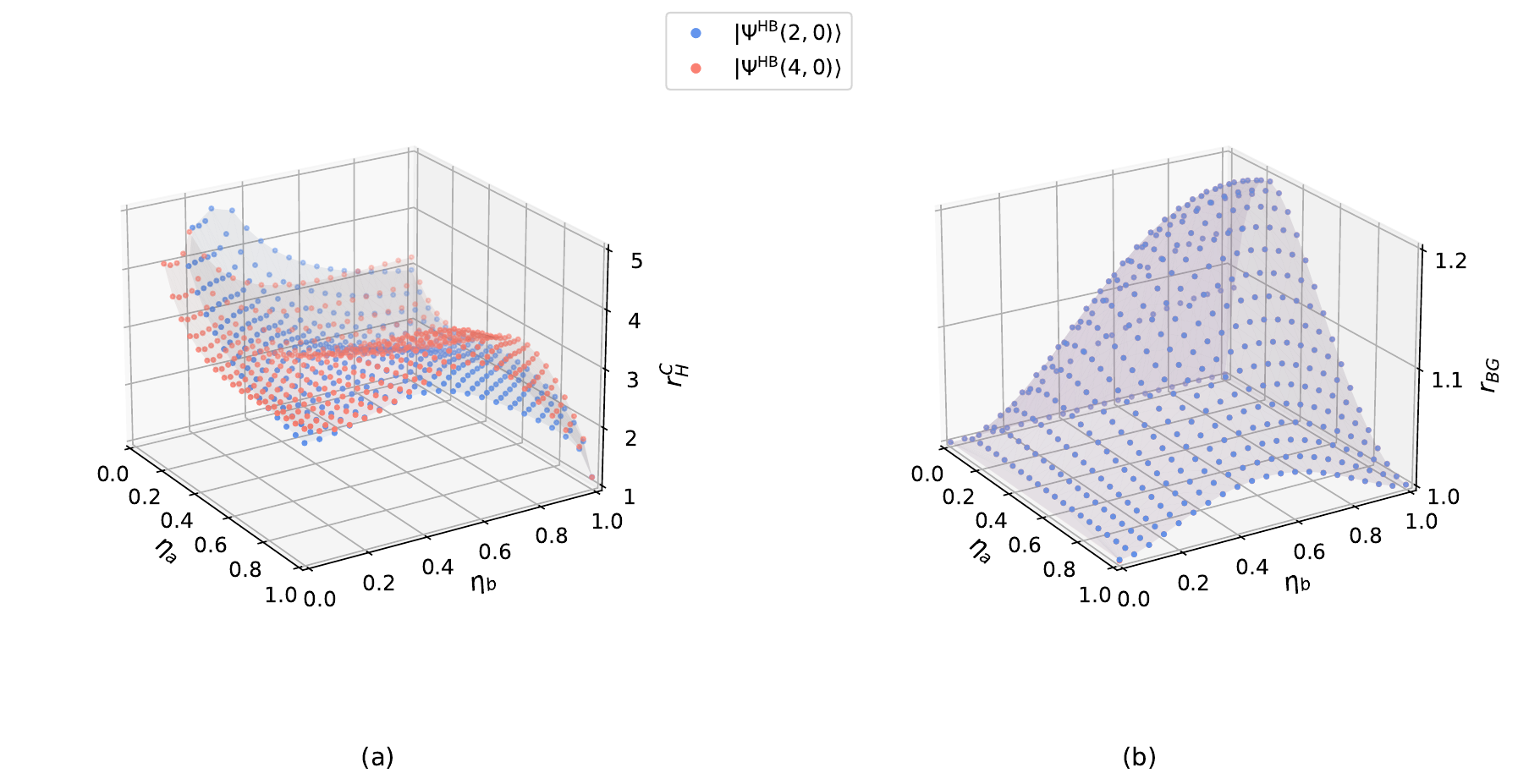}
\caption{\small{\textit{Joint-estimation of phase and loss:} Three-dimensional plots of double homodyne measurement compatibility $r^C_H$ (panel (a)) and fundamental compatibility $r_{BG}$ (panel (b)) seen as a function of the losses, $\eta_a$ and $\eta_b$, in each arm with equal parameter weights i.e., at $y=0.5$. The probe state considered here is the gHB state $\ket{\Psi_{\mathrm{gHB}}(N,0)}$ with $N=2$ (blue dots) and $N=4$ (red dots) which create smooth surfaces. Note that the points closer to low values of losses exhibit high measurement and fundamental compatibilities. Also, note that the fundamental compatibility does not vary with respect to the chosen values of $N$ for almost all pairs of values of losses resulting in overlapping points.}}
\label{fig:dhd_3D}
\end{figure*}

\begin{figure*}
\centering
\includegraphics[scale=0.55]{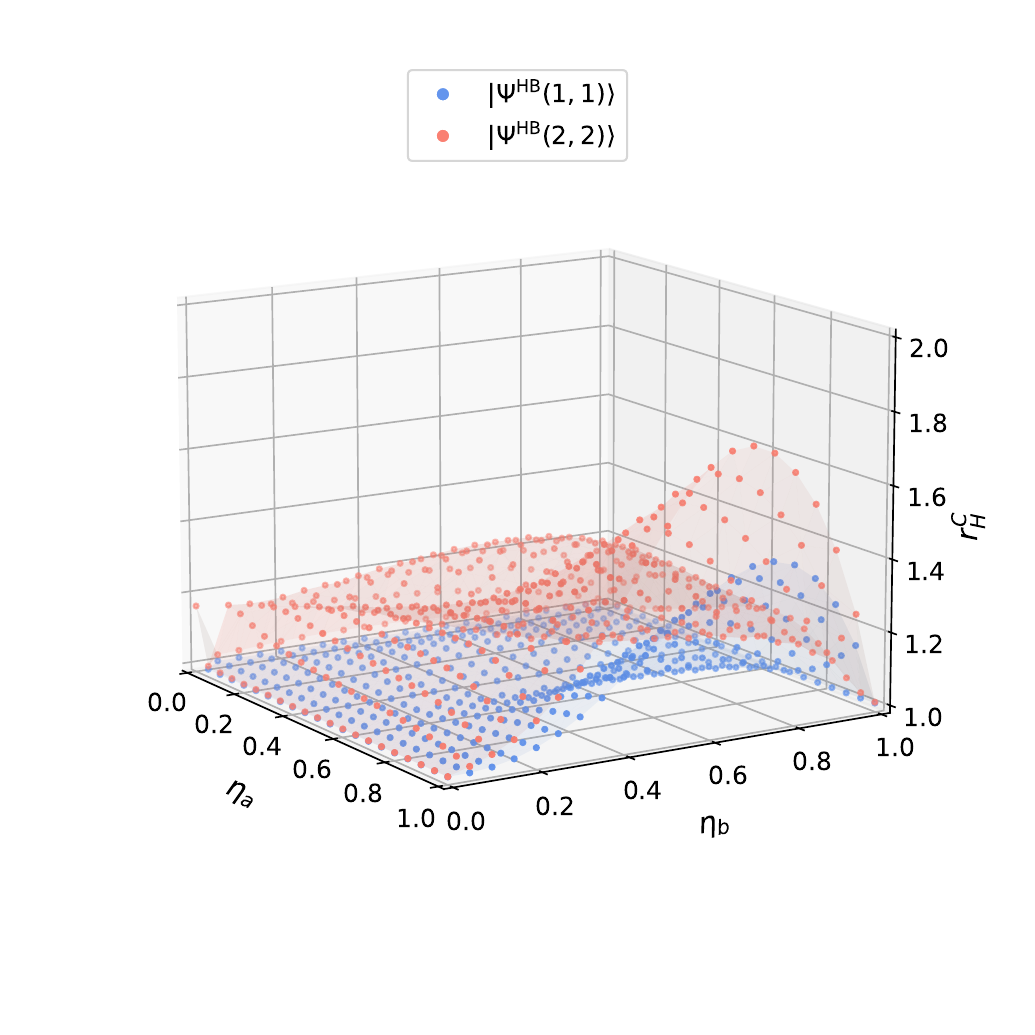}
\caption{\small{\textit{Joint-estimation of phase and loss for photon counting:} Three-dimensional plots of measurement compatibility $r^C_H$ versus losses at equal weights. The probe state considered here are the HB states: $\ket{\Psi_{\mathrm{gHB}}(1,1)}$ and (blue dots) and $\ket{\Psi_{\mathrm{gHB}}(2,2)}$ (red dots) which create smooth surfaces. Note that for the combinations of $\eta_a$ and $\eta_b$ plotted here, the compatibility remains relatively high.}}
\label{fig:pn_3D}
\end{figure*}

An analysis at different levels of losses $\eta_a$ and $\eta_b$ is reported in Fig.~\ref{fig:dhd_3D} for the double homodyne measurement, illustrating its performance in terms of compatibility of different states for equal weights for the two parameters, $y=0.5$. For moderate loss, measurement compatibility (panel \ref{fig:dhd_3D}(a)) only shows a weak dependence on $\eta_a$ and $\eta_b$, while a decrease in transmission entails a decrease of the compatibility as $N$ grows. Inspection of the fundamental compatibility $r_{BG}$ (panel \ref{fig:dhd_3D}(b)) shows that, in the central region, compatibility is not assured for intermediate loss: this makes it relatively easier to approach the Holevo limit with a realistic measurement. A similar analysis for photon counting with respect to different values of losses is illustrated in Fig.~\ref{fig:pn_3D}. One can infer that, overall, this choice of the measurement offers better compatibility than homodyne, as its value remains below 1.8. Photon counting seems more effective in achieving the limit of HCRB in this multiparameter example.

\newpage
\subsection{Joint-estimation of phase and phase diffusion}\label{sec:pd}

\begin{figure*}
\centering
\includegraphics[scale=0.5]{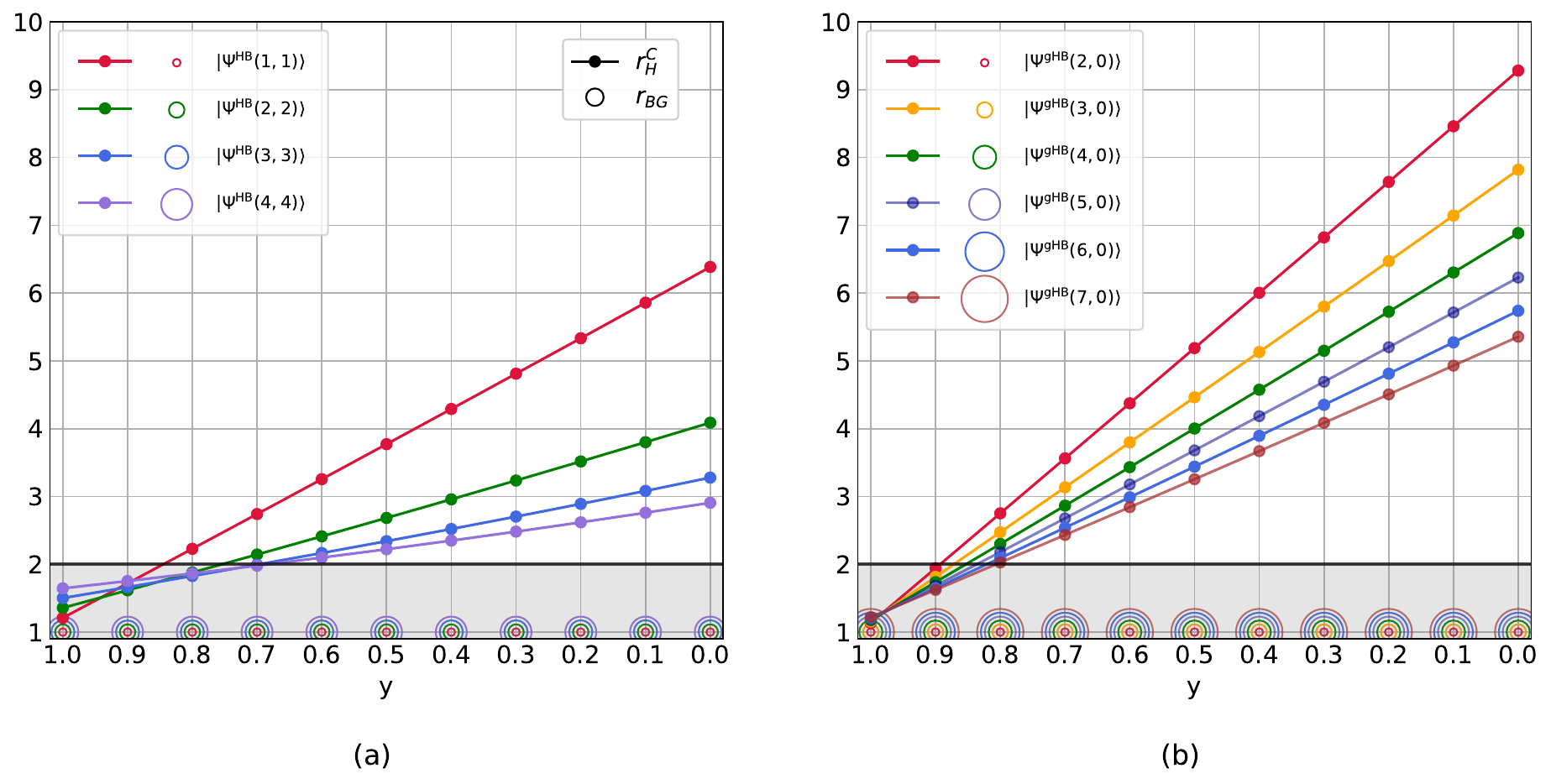}
\caption{\small{Plot of measurement compatibility measure $r^C_H$ and fundamental compatibility measure $r_{BG}$ versus the weights $y$ for the joint-estimation of phase and phase diffusion in the MZI with equal losses in both arms. HB states with $N=2,4,6,8$ (panel (a)) and gHB states with $N=2,3,4,5,6,7$ (panel (b)), and double homodyne measurement are considered. The curves are plotted at $\eta=0.999$ and $\Delta=0.1$.}}
\label{fig:dhd_pd}
\end{figure*}

\begin{figure*}
\centering
\includegraphics[scale=0.5]{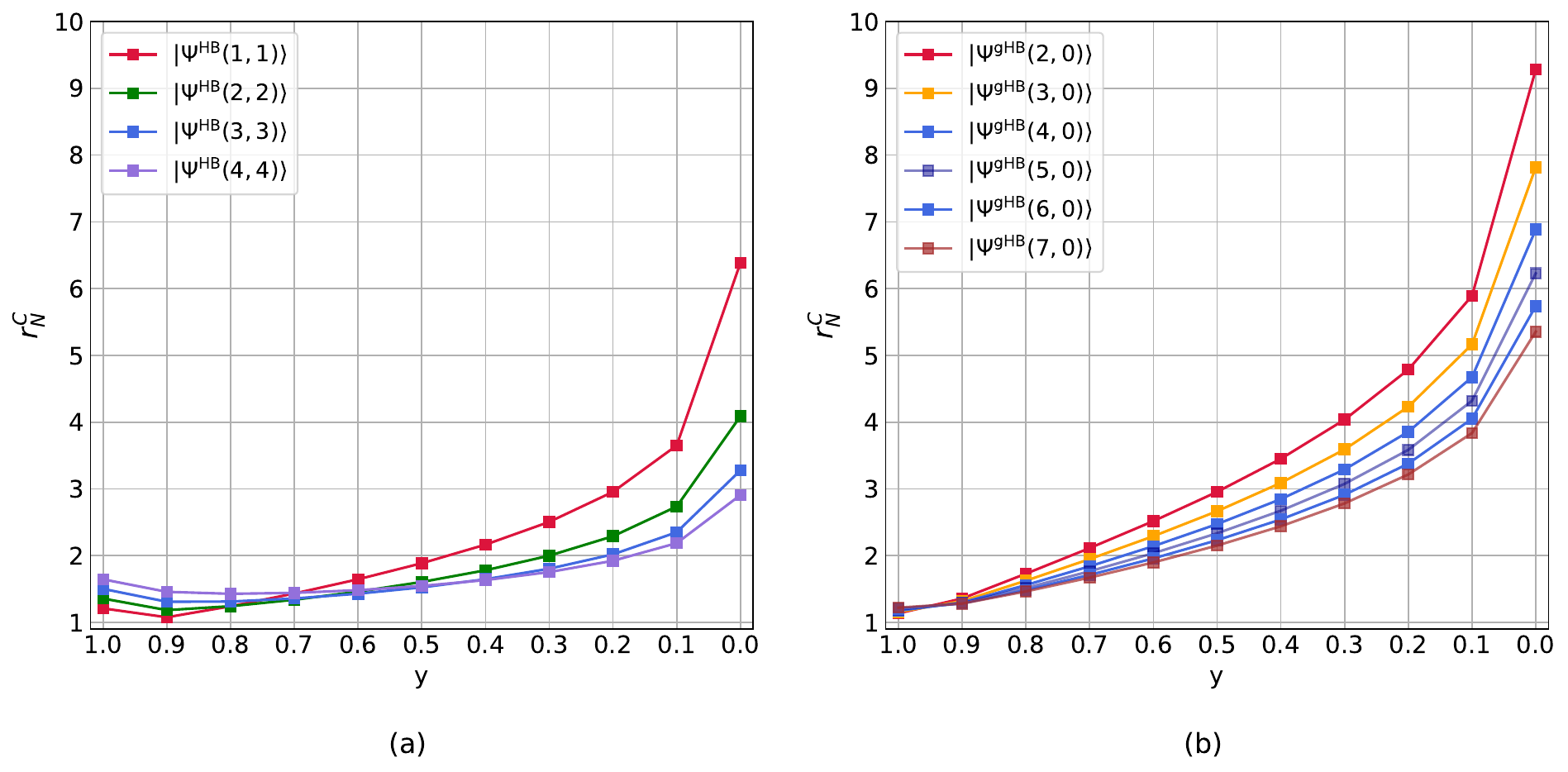}
\caption{\small{Plot of measurement compatibility measure $r^C_N$ (with respect to NHCRB) versus the weights $y$ for the joint-estimation of phase and phase diffusion in the MZI with equal losses in both arms. HB states with $N=2,4,6,8$ (panel (a)) and gHB states with $N=2,3,4,5,6,7$ (panel (b)), and double homodyne measurement are considered. The curves are plotted at $\eta=0.999$ and $\Delta=0.1$.}}
\label{fig:dhd_pd_nhb}
\end{figure*}

In last example, we consider the joint estimation of phase and phase diffusion while assuming known amount of losses in both arms.  For simplicity, the output probe state $\rho^{\mathrm{gHB}}_{\phi,\Delta}$ is obtained from Eq.\,\ref{eqn:outputprobe} by setting equal amount of losses in both arms, $\eta_a=\eta_b=\eta$, yielding\\
$\rho^{\mathrm{gHB}}_{\phi,\Delta}(n,N-n)=\newline\sum_{k=0}^{N}\sum_{l=0}^{N-k}\sum_{p,q=k}^{N-l}\mathcal{C}^{N}_{\phi,\Delta}(n,p,q)\\
\ket{p-k,N-p-l}\bra{q-k,N-q-l}$.

In Fig.\,\ref{fig:dhd_pd}, panel (a) represents the plots of $r^C_H$ and $r_{BG}$ versus $y$ for HB states with $N=2,4,6,8$ and panel (b) represents the same plots for gHB states with $N=2,3,4,5,6,7$. We set $\Delta=0.1$ and $\eta=0.999$ (99.9\% transmission loss in both arms).

Firstly, we report a recognisable behavior of $r_{BG}$: since phase and phase diffusion correspond to observables that are weakly compatible i.e.,\\ $\Tr(\rho^{\mathrm{gHB}}_{\phi,\Delta}[L_{\phi},L_{\Delta}])=0$~\cite{vidrighin2014joint,szczykulska2017reaching,jayakumar2024quantum}, the system reaches maximal fundamental compatibility i.e., $r_{BG}=1$ for all values of $y$ in both panels. As for the behavior of $r^C_H$, in panel \ref{fig:dhd_pd} (a), we find an inversion of the ordering as for the case of loss in Fig.\,\ref{fig:dhd_ba}, but this time with the opposite trends.  The curves intersect around $y=0.9$ due to which, for $y\leq0.9$, the overall magnitude of $r^C_H$ increases as $N$ increases, whereas for $y>0.9$, the magnitude decreases as $N$ increases. In panel \ref{fig:dhd_pd}(b), most features observed in panel \ref{fig:dhd_pd}(a) remain the same, except that the dependence of $r^C_H$ on $N$ at $y=1$ is much less pronounced.

For the photon counting measurement, we find that the FIM is singular for HB states, while for gHB states it reaches large values in general, and singularity may occur. We could attribute this to the fact that intensity measurement at the outputs of the MZI cannot distinguish a decrease in contrast due to the shift of the fringes (associated to $\phi$) or to an overall reduction of visibility (associated to $\Delta$). An auxiliary phase shift may be introduced in order to resolve such an ambiguity~\cite{roccia2018multiparameter}.

The weak compatibility in this multiparameter example determines that an advantage may exist in the use of collective measurements, as captured by the HCRB. It is thus meaningful to focus on optimal separable strategies, identified by the NHCRB, thus using the associated compatibility $r^N_H$. We find that our example is characterized by the presence of a gap: $1\leq r^N_H \leq 2$. Notice that, since we deal with a two-parameter estimation $(p=2)$, the measure cannot exceed 2 as mentioned in Section~\ref{sec:foms}. As a result, the quantity $r^C_N$ earns its relevance here and is plotted in Fig.~\ref{fig:dhd_pd_nhb} against the parameter weights, for the same set of probe states and parameter values of $\eta$ and $\Delta$ as in Fig.~\ref{fig:dhd_pd}. We desume that $r^C_N$ paints a more optimistic picture of our measurements, with higher compatibility values i.e., $r^C_N\leq 2$ achieved when $0.5\leq y \leq 1$ in panel \ref{fig:dhd_pd_nhb} (a), and $r^C_N\leq 2.2$ when $0.7\leq y\leq 1$ in panel \ref{fig:dhd_pd_nhb}(b), for all values of $N$ considered.

\section{Discussion}

A closer inspection of the performance of the double homodyne measurement under varying losses in each mode, $\eta_a$ and $\eta_b$, with equal parameter weights (see Fig.~\ref{fig:dhd_3D}(a)), reveals important insights into the practical relevance of our results. For the state $\ket{\Psi_{\text{gHB}}(2,0)}$, we find that the measurement compatibility, $r^C_H$, remains below 2.2 for a number of coordinates $(\eta_a,\eta_b)$ within the ranges: $0.8\leq \eta_a\leq1$ and $0.8\leq \eta_b\leq1$. For the state $\ket{\Psi_{\text{gHB}}(4,0)}$, the region corresponding to the same upper bound of $r^C_H$ occurs within the ranges: $0.9\leq \eta_a\leq1$ and $0.9\leq \eta_b\leq1$. Although, these regions may appear rather far from the ideal limit, they still offer relatively high compatibility in practice. We base this on the fact that the double homodyne measurement is viewed as an outstanding solution in metrology for its high efficiency and high noise rejection. For the case of phase diffusion, instead, in comparison with Fig.~\ref{fig:dhd_pd}, the results in Fig.~\ref{fig:dhd_pd_nhb} shed a more optimistic light when the measurement is benchmarked against the best-performing separable measurement characterized by the NHCRB. The relevance of $r^C_H$ becomes even more clear when considered alongside the fundamental compatibility, $r_{BG}$. Owing to the fact that the value of $r_{BG}$ remains consistently below 1.05 not only in the ranges: $0.8\leq \eta_a\leq1$ and $0.8\leq \eta_b\leq1$ but also in the high loss regions (Fig.~\ref{fig:dhd_3D}(b)), the double homodyne measurement is closer to optimal information extraction.

In Figs.\ref{fig:dhd_oa}-\ref{fig:dhd_pd}, we have inspected the behavior of measurement and fundamental compatibility under lossy scenarios. To further elucidate the connection between compatibility and losses in the system, we consider an ideal scenario of minimal losses i.e., $\eta_a=\eta_b=0.999$, $\Delta=0$ in Eq.\,\ref{eqn:outputprobe} while estimating phase and loss in one arm jointly. Considering HB and gHB states with $N=2,4,6$,  numerical results show that $r_{BG}\approx1$ for all values of $y$, indicating maximal fundamental compatibility. As for the measurement compatibility, we find that the $r^C_H\approx1$ at $y=1$ reaching maximal measurement compatibility due to the optimality of the double homodyne measurement in estimating phase~\cite{vidrighin2014joint}. Moreover, $r^C_H$ remains below 1.5 for $y\geq0.5$ for all the probe states considered (see Appendix\,\ref{app:minloss}), hence  the double homodyne measurement closely attains the HCRB even with individual copies. This can be understood by considering that $r^C_H$ may be approximated as
\begin{equation}\label{eqn:mceqn}
r^C_H\approx\frac{2y(F_{C\,\phi})^{-1}+2(1-y)(F_{C\,\eta})^{-1}}{2y(F_{Q\,\phi})^{-1}+2(1-y)(F_{Q\,\eta})^{-1}}
\end{equation}

since $C_H\approx C_Q$ and $F_{C,Q\,i,j}=0$ for $i\neq j$ . For low loss, the conditions $F_{C\,\eta}\gg F_{C\,\phi}$ and $F_{Q\,\eta}\gg F_{Q\,\phi}$ are satisfied, implying that their contribution to the measurement compatibility is small around $y=1$. As a result, the dominant contribution comes from $F_{C\,\phi}$ and $F_{Q\,\phi}$, and the ensuing optimality of the measurement keeps the ratio $r^C_H$ fairly constant and closer to 1. This behavior can also be seen in Fig.\,\ref{fig:dhd_3D}, in the low-loss regime with $y = 0.5$.

The collective measurements required to attain the HCRB are experimentally challenging to implement. Therefore, with the view of practical feasibility, one must turn towards finite-copy separable measurements, although this comes at a cost of reduced precision. Firstly, we mention that in the case of pure state models, the HCRB is attainable with optimal measurements already at a single-copy level~\cite{matsumoto2002new}. Secondly, for more general models, the NHCRB is attainable with separable measurement on finite number of identical copies of the probe state in many cases. These include the estimation of qubit rotations in the presence of phase damping using a two qubit probe and more interestingly, the joint-estimation of phase and loss in one arm~\cite{conlon2021efficient}. However, it must be noted that the NHCRB is not always tight—for instance, in the model described in Example~$A$ of~\cite{conlon2025role}. In this case, although $C_N = C_Q$, the presence of fundamental incompatibility renders the NHCRB unattainable with any separable measurement. A tight, ultimate precision bound for separable measurements was formulated by Hayashi~\cite{hayashi1997linear}, and more recently, a strict gap between this tight bound and the NHCRB was demonstrated using conic programming~\cite{hayashi2023tight}, further cautioning that there exist cases where the NHCRB is not attainable.

For the joint-estimation of phase and loss in one arm, a single-copy optimal measurement obtained from the SLD for phase with random single-photon input states of the form:\\ $\ket{\psi}=c_0\ket{0,1}+c_1\ket{1,0}$ has been demonstrated to fully attain the HCRB for $\eta_a\geq0.5$~\cite{albarelli2019evaluating}. Furthermore in~\cite{albarelli2019evaluating}, it has been numerically demonstrated that there exists certain values of $N$ and $\eta_a$ at which this measurement attains the HCRB. This inference is in agreement with our numerical result $r^N_H = 1$. We can conjecture that this is due to the fact that the output probe, in this case, is expressed as a direct sum of pure states corresponding to the number of lost photons and there exists a separable measurement with a direct-sum structure achieving the HCRB~\cite{matsumoto2002new}. However, in the case of phase diffusion (Section\,\ref{sec:pd}), the loss of direct sum structure may signal the impossibility to attain the HCRB at a single-copy level. Furthermore, it has been shown that if the HCRB cannot be attained at a single-copy level, it can never be attained with any finite number of copies, which makes the single-copy attainability a fundamental one ~\cite{conlon2208gap}. Of course, collective measurements on infinitely many copies are known to saturate the HCRB, but they are resource inefficient. Nevertheless as mentioned earlier, with finite resources and separable measurements, one can still attain the NHCRB.

We end the discussion with remarks on the following features of the single parameter bounds for the estimation of phase and loss. Our numerics show that, independently on $N$, the QFI for loss in HB and gHB probe states is always larger than that for phase. At the measurement level, however, this condition is not necessarily met by the FI matrix of the double homodyne measurement. If one considers the optimal states for phase and loss estimation, there is a crossing of the QFIs for loss and phase exactly at $\eta=0.5$. As a result, one can infer that in the regime of high loss i.e., $\eta<0.5$, it is harder to estimate phase than loss, whereas in the regime of low loss i.e., $\eta>0.5$, it is harder to estimate loss than phase.

\section{Conclusion}

In this work, we have investigated the measurement compatibility in quantum interferometry, considering the joint estimation of phase along with (i) loss in one arm, (ii) loss in both arms, and (iii) phase diffusion in a MZI. These examples are motivated by their practical relevance, but also by conceptual interest: the strong fundamental incompatibility between phase and loss and the weak fundamental compatibility between phase and phase diffusion. We have differentiated the behavior of the measurement compatibility with the fundamental compatibility that is intrinsic to the estimation problem at hand. As our primary aim, we have examined the performance with respect to the HCRB of resource-efficient and practical separable measurements, as opposed to the ideal but experimentally expensive collective measurements. Our results highlight how keeping loss low has the benefit of improving the compatibility of multiparameter estimation, in addition to the usual advantage of increasing the level of practically useful resources.

Multiparameter estimation is by necessity a multifaceted problem, and this entails keeping the attention on all relevant aspects simultaneously. The measurement compatibility $r^C_H$ does not intend to be an all-encompassing figure, but adds to the quantum estimation toolbox when it comes to considering this one aspect. We thus anticipate it will become commonplace in the analysis of quantum sensing schemes. 

\begin{acknowledgments}
We thank Francesco Albarelli for stimulating discussion.
J. J. and M. S. were supported by the National Science Centre `Sonata Bis' Project No. 2019/34/E/ST2/00273 and the Foundation for Polish Science `First Team' Project No. POIR.04.04.00-00-220E/16-00 (originally, FIRST TEAM/2016-2/17). M. S. was supported by the European Union's Horizon 2020 research and innovation programme under the Marie Skłodowska-Curie project `AppQInfo' No. 956071, and the QuantERA II Programme that has received funding from the European Union's Horizon 2020 research and innovation programme under Grant Agreement No. 101017733, project `PhoMemtor' No. 2021/03/Y/ST2/00177.
M.B. acknowledges support from the PRIN project PRIN22-RISQUE-2022T25TR3 and MUR Dipartimento di Eccellenza 2023-2027 of the Italian Ministry of University.
\end{acknowledgments}

\section*{Author Contributions}

J.J., M.B., and M.S. jointly conceptualized the work. J.J. performed the numerical computations and prepared all figures. The manuscript was written by J.J. and M.B., with critical revisions and final approval provided by M.S., who also supervised the project.

\section{Appendix}
\subsection{The Holevo Cram\'er-Rao bound}\label{app:hcrb}

In Eq.~\ref{eqn:hcrb1}, the matrices $X_i$ are related to estimators and POVM for a measurement $M$ given by 

\begin{equation}\label{eqn:x_mat}
X_i=\int dx (\tilde{\theta_i}-\theta_i)\Pi_x.
\end{equation}

Furthermore, considering a general probe state $\rho_{\vec{\theta}}$ and setting $\nu=1$, the minimization over $V$ can be performed analytically which results in the following reformulated version of the HCRB~\cite{albarelli2020perspective}.

\begin{equation}\label{eqn:reform}
\begin{aligned}
    C_H(\rho_{\vec{\theta}},W,\vec{\theta}) 
    &= \min\limits_{\vec{X}} \Big[ \Tr(W\mathrm{Re}Z(\vec{X})) \\
    &\quad +\lVert \sqrt{W}\mathrm{Im}Z(\vec{X})\sqrt{W}\rVert_{1} \\
    &\quad \mid \Tr(\nabla\rho_{\vec{\theta}}\vec{X}^{\mathrm{T}}) = \mathds{1} \Big]
\end{aligned}
\end{equation}

where $\lVert A \rVert_1=\Tr(\sqrt{AA^{\dagger}})$ is the trace norm. This reformulation is more informative in the sense that one can prove that the minimization involving $\mathrm{Re}(Z(\vec{X}))$ alone gives exactly the QCRB~\cite{ragy2016compatibility} such that

\begin{equation} \min\limits_{\vec{X}}\Tr(W\mathrm{Re}Z(\vec{X}))=\Tr(WF^{-1}_Q)=C_Q.
\end{equation}

After performing the full minimization in Eq.\,\ref{eqn:reform} i.e., involving both $\mathrm{Re}(Z(\vec{X}))$ and $\mathrm{Im}(Z(\vec{X}))$, the contribution due to the fundamental incompatibility is revealed. Thus, it is evident that the imaginary part of $Z(\vec{X})$ contains information about the fundamental incompatibility.

For an explicit quantitative understanding of fundamental compatibility, it is useful to consider the closed-form upper bound to $C_H$ which reads

\begin{equation}
    C_H \leq C_Q+ \lVert \sqrt{W} F^{-1}_Q I F^{-1}_Q\sqrt{W}\rVert_1=\overline{C}_H,
\end{equation}

where $I_{i,j}=\frac{1}{2}\Tr(\rho_{\vec{\theta}}[L_{\theta_i},L_{\theta_j}])$. For the D-invariant models, these bounds coincide i.e.,\\ $C_H=\overline{C}_H$.

In general, a necessary and sufficient condition for the equivalence of HCRB and QCRB is given by

\begin{equation}
    \Tr\big(\rho_{\vec{\theta}} [L_{\theta_i}, L_{\theta_j}]\big) = 0, \quad \forall \, \theta_i, \theta_j \in \Theta, \, i \neq j
\end{equation}

This condition is known as the commutation condition \cite{matsumoto2002new,vaneph2013quantum,ragy2016compatibility,suzuki2016explicit} indicating weak fundamental compatibility. 

\subsection{The Nagaoka-Hayashi Cram\'er-Rao bound}\label{app:nhcrb}

We start by noting that the covariance matrix $\Sigma$ may be rewritten as

\begin{equation}
    \Sigma_{i,j}=\int dx \Tr(\rho_{\vec{\theta}}\Pi_x)(\tilde{\theta}_i-\theta_i)(\tilde{\theta}_j-\theta_j)
\end{equation}

\begin{equation}
    =\Tr(\rho_{\vec{\theta}}\int dx(\tilde{\theta}_i-\theta_i)\Pi_x(\tilde{\theta}_j-\theta_j))=\Tr(\rho_{\vec{\theta}}\mathbb{L}_{i,j})
\end{equation}

where we introduce the matrix $\mathbb{L}$ with elements $\mathbb{L}_{i,j}=\int dx(\tilde{\theta}_i-\theta_i)\Pi_x(\tilde{\theta}_j-\theta_j))$. Furthermore, we explicitly denote $\Sigma(\rho_{\vec{\theta}},\mathbb{L})=\Tr(\rho_{\vec{\theta}}\mathbb{L})$, a real symmetric matrix belonging to $\mathcal{S}^{\mathbb{R}}(\Theta)$.

Given these considerations, the NHCRB may be obtained as the solution of the following convex optimization problem

\begin{equation}\label{eqn:nhcrb2}
\begin{aligned}
    C_N(\rho_{\vec{\theta}},W,\vec{\theta}) &=  \min\limits_{\mathbb{L},\vec{X}} \Big[ \Tr(W\Sigma(\rho_{\vec{\theta}},\mathbb{L})) \\
    & \quad \begin{cases}
             \mathbb{L}\geq \vec{X}\vec{X}^{\mathrm{T}}\\      \Tr(\nabla\rho_{\vec{\theta}}\vec{X}^{\mathrm{T}})=\mathds{1}
    \end{cases}\Bigg]
\end{aligned}
\end{equation}

Eq.~\ref{eqn:nhcrb2} is cast as an SDP and solved numerically~\cite{conlon2021efficient,zhang2022quanestimation}. 

In order to see the relation between the formulation of NHCRB and the HCRB, the latter may be rewritten in the following form

\begin{equation}
\begin{aligned}
    C_H(\rho_{\vec{\theta}},W,\vec{\theta}) &=  \min\limits_{\mathbb{L},\vec{X}} \Big[ \Tr(W \Sigma(\rho_{\vec{\theta}},\mathbb{L})) \\
    & \quad \begin{cases}
             \Sigma(\rho_{\vec{\theta}},\mathbb{L})\geq \Tr(\rho_{\vec{\theta}}\vec{X}\vec{X}^{\mathrm{T}})\\      \Tr(\nabla\rho_{\vec{\theta}}\vec{X}^{\mathrm{T}})=\mathds{1}
    \end{cases}\Bigg]
\end{aligned}
\end{equation}

Comparing with Eq.~\ref{eqn:hcrb1}, we identify $V=\Sigma(\rho_{\vec{\theta}},\mathbb{L})$ and $Z(\vec{X})=\Tr(\rho_{\vec{\theta}}\vec{X}\vec{X}^{\mathrm{T}})$. In particular, for projective measurements i.e., when $\Pi_i\Pi_j=\delta_{i,j}\Pi_i$, one can show $Z(\vec{X})=\Sigma(\rho_{\vec{\theta}},\mathbb{L})$~\cite{demkowicz2020multi}. Therefore, in the HCRB, we simply optimize over all the covariance matrices $\Sigma$ which causes the information about the optimal separable measurement to be lost (which is contained in $\mathbb{L}$ and $\vec{X}$). However, in the NHCRB, we optimize particularly over $\mathbb{L}$, and hence the information about the optimal separable measurement is preserved, yielding more information. In this sense, the NHCRB is more informative than the HCRB resulting in $C_N\geq C_H$.

Lastly, in cases where the NHCRB is attainable, the optimal separable measurement is estimated from the solution $X_i$ of the NHCRB and Eq.~\ref{eqn:x_mat}.

\subsection{The double homodyne measurement}\label{app:dhd}

In this section, we provide a concise summary of the double homodyne measurement which involves two homodyne measurements of the output field quadratures $\hat{x}$ and $\hat{p}$. This is based on the description provided in the supplementary material of~\cite{vidrighin2014joint}. As a result, it is a continuous variable measurement complementary to the photon counting measurements which involves the measurement of discrete variables. The quadrature operator is parametrized by the phase $\varphi$ as $\hat{x}(\varphi)=\frac{1}{2}(\hat{a}e^{-i\varphi}+\hat{a}^{\dagger}e^{i\varphi})$, and it is controlled by the phase $\varphi_l$ of the local oscillator, a strong coherent state that interferes with one of the modes of the output state. The eigenstates of the quadrature operator can be written in the photon number basis as follows~\cite{welsch1999ii}

\begin{equation}\label{eqn:quad}
    \ket{x,\varphi}=\pi^{-1/4}e^{-x^2/2}\sum_{n=0}^{\infty}\frac{e^{in\varphi}}{\sqrt{2^nn!}}\mathrm{H}_{n}(x)\ket{n},
\end{equation}

where $\mathrm{H}_{n}(x)$ are the Hermite polynomials. It is evident that the measurement of the output state involves projecting it onto coherent state-like projectors, thereby, constituting a Gaussian measurement. We integrate the output beamsplitter of the MZI as a part of the double homodyne measurement scheme. Therefore, for the measurement of $\hat{x}$ ($\hat{x}(0)$) and $\hat{p}$ ($\hat{x}(\pi/2)$) operators with eigenvalues $x$ and $p$ respectively, we may write \\$\ket{\nu(x,p)}=\mathcal{U}_{\mathrm{BS}}^{(\mathrm{o})}\ket{x,p}$, where $\mathcal{U}^{\mathrm{(o)}}_{\mathrm{BS}}$ is the output balanced beamsplitter operation, i.e., with transmissivity $T = 1/2$ and a phase shift of $\pi/2$ between the reflected and transmitted beams. The corresponding POVM reads\\ $\Pi_{x,p}=\ket{\nu(x,p)}\bra{\nu(x,p)}$. Defining the polar coordinates $(r,\chi)$, where $r=\sqrt{x^2+p^2}$ and $\chi=\tan^{-1}\big(p/x\big)$, and making use of Eq.~\ref{eqn:quad}, $\ket{\nu(x,p)}$ takes the following explicit form

\begin{equation}
    \ket{\nu^{N}_{kl}(r,\chi)}=\sum_{n=k}^{N-l} g_{n,N-n}(r)e^{i(2n-N)\chi}\ket{n-k,N-n-l}
\end{equation}

where $g_{n,N-n}(r)=\\ \begin{cases} 
\begin{aligned}
\sqrt{\frac{(n-k)!}{(N-n-l)!\pi}}e^{-r^2/2}r^{N-2n+k-l} \\
\times(-1)^{N-2n+k-l}\mathrm{L}^{(N-2n+k-l)}_{n-k}(r^2)
\end{aligned} & \scriptstyle \text{if } N - 2n + k - l > 0 \\[1em]
\begin{aligned}
\sqrt{\frac{(N-n-l)!}{(n-k)!\pi}}e^{-r^2/2}r^{2n-N-k+l} \\
\times \mathrm{L}^{(2n-N-k+l)}_{N-n-l}(r^2)
\end{aligned} & \scriptstyle \text{if } N - 2n + k - l < 0 \\[1em]
\frac{e^{-r^2/2}}{\sqrt{\pi}}\mathrm{L}_{n-k}(r^2) & \scriptstyle \text{if } N - 2n + k - l = 0
\end{cases}$ with $k$ and $l$ lost photons included to reflect the lossy state on which the measurement is made, $\mathrm{L}^{(\alpha)}_{n}(r^2)$ are the generalized Laguerre polynomials, and $g_{n,N-n}(r)$ must satisfy $\int_{0}^{\infty}dr[g_{n,N-n}(r)]^2r=1/2\pi$.

The POVM $\Pi^N_{kl}(r,\chi)=\ket{\nu^{N}_{kl}(r,\chi)}\bra{\nu^{N}_{kl}(r,\chi)}$ satisfies \\$\int_{\chi=0}^{2\pi}\int_{r=0}^{\infty}d\chi dr\Pi^N_{kl}(r,\chi)r=\mathds{1}_{kl}$ i.e., integrating to identity matrix in the subspaces when $k$ and $l$ photons are lost and $\Pi^N_{kl}(r,\chi)\geq0$. The corresponding probability distribution with respect to the output state reads\\ $p_{\vec{\theta}}(r,\chi)=\Tr(\rho^{\mathrm{gHB}}_{\vec{\theta}}\Pi^N_{kl}(r,\chi))$ ; $\vec{\theta}=(\phi,\eta_a,\Delta)^\top$.

\subsection{Measurement compatibility with respect to the weights under varying loss on the reference arm}\label{app:refarm}

\begin{figure*}[t]
\centering
\includegraphics[scale=0.48]{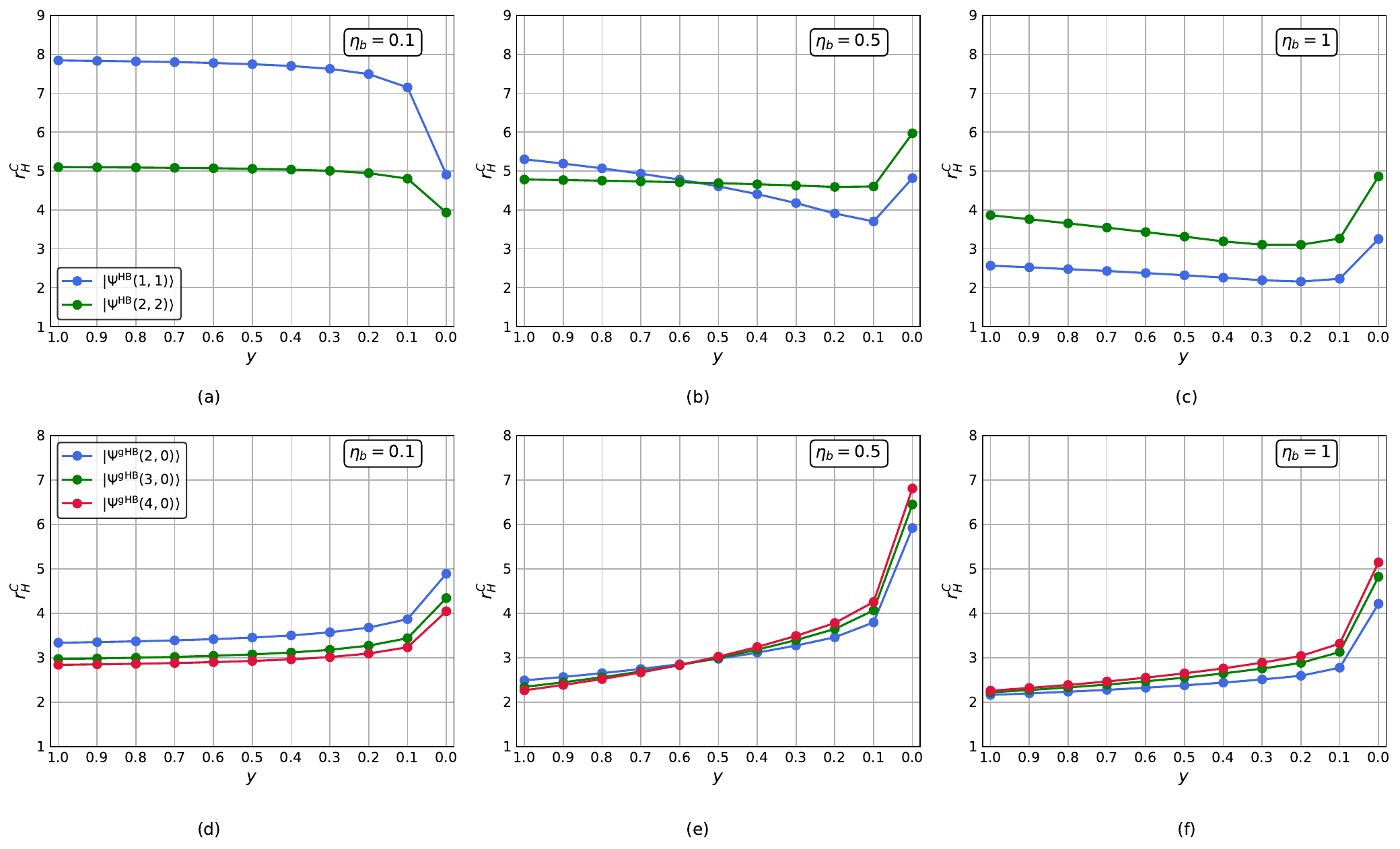}
\caption{\small{\textit{Top row:} Plot of measurement compatibility measure $r^C_H$ versus the weights $y$ for the joint-estimation of phase and loss considering HB states with $N=2,4$ at different values of loss on reference arm, namely, $\eta_b=0.1$, $\eta_b=0.5$, and $\eta_b=1$. \textit{Bottom row:} The same plots for gHB states with $N=2,3,4$. The double homodyne measurement is considered and the curves in all the panels are plotted at $\eta_a=0.5$ and $\Delta=0$.}}
\label{fig:dhd_refarm}
\end{figure*}

In this section, we demonstrate the behavior of measurement compatibility with respect to the weights for the double homodyne measurement as one varies the loss on the reference arm $\eta_b$ by keeping $\eta_a$ at a fixed value. In Fig.\,\ref{fig:dhd_refarm}, we have obtained the plots for gHB states with $N=2,3,4$ and HB states with $N=2,4$. In the regime of high losses on the reference arm i.e., $\eta_b=0.1$, the overall value of $r^C_H$ decreases as $N$ increases. However, with $\eta_b=0.5$, there is a crossing at $y=0.6$. When the loss in the reference arm is low, the overall value of $r^C_H$ increases as $N$ increases. We infer that these features are an artifact of the measurement since they remain qualitatively the same for the class of probe states considered here.

\subsection{Compatibility with respect to the weights under minimal losses}\label{app:minloss}

\begin{figure*}[t]
\centering
\includegraphics[scale=0.7]{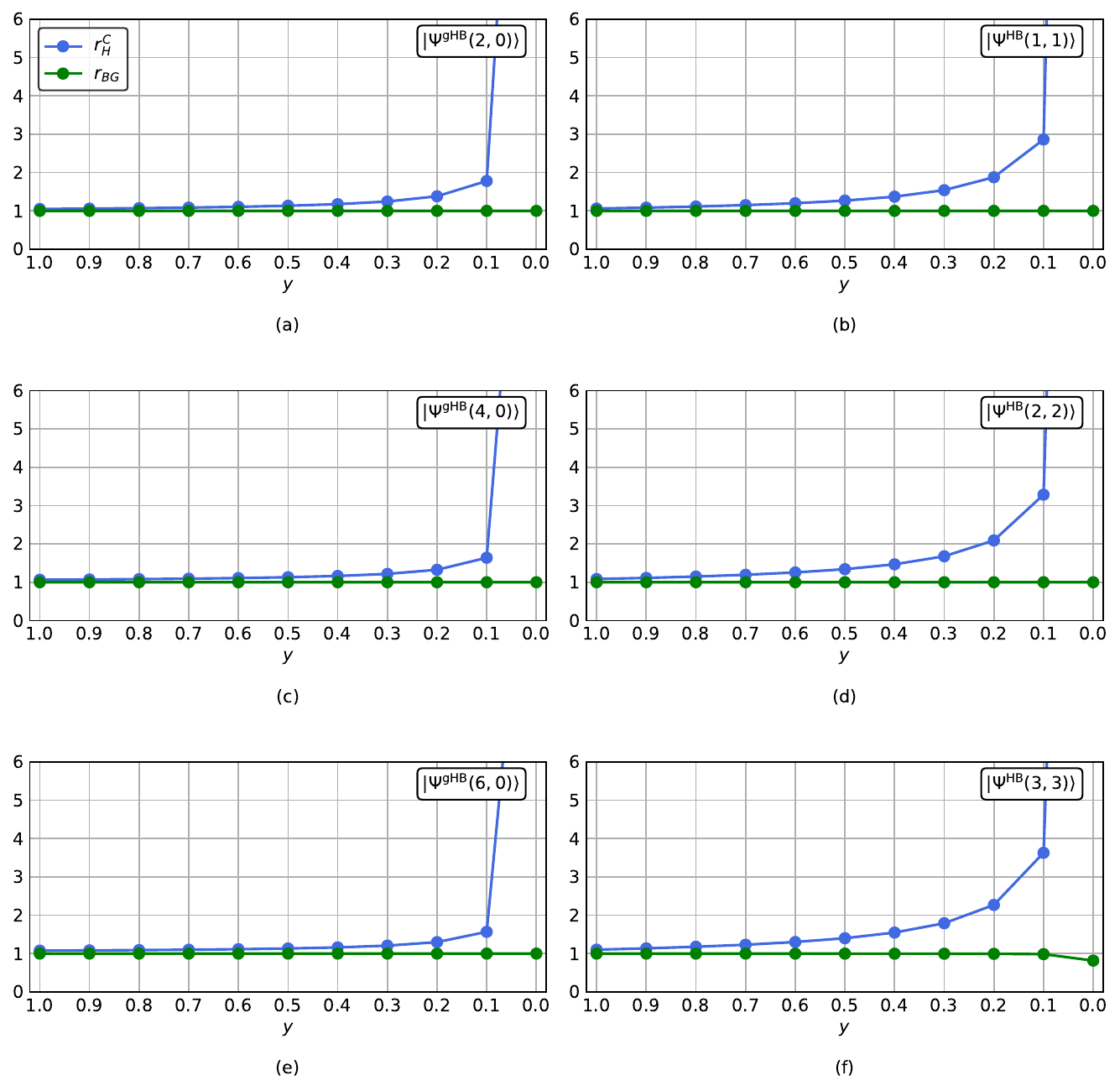}
\caption{\small{Plot of measurement compatibility $r^C_H$ and fundamental compatibility $r_{BG}$ versus the weights $y$ for the joint-estimation of phase and loss with $\eta_a=\eta_b=0.999$ and $\Delta=0$ (minimal losses). gHB and HB states with $N=2,4,6$ and double homodyne measurement are considered here. Note that for $y\geq0.5$, $r^C_H$ remains below 1.5 implying that the HCRB is closely attained by the double homodyne measurement.}}
\label{fig:dhd_minloss}
\end{figure*}

In this section, in Fig.\,\ref{fig:dhd_minloss}, we depict the behavior of both measurement and fundamental compatibilities with respect to the weights under minimal losses in both arms i.e., $\eta_a=\eta_b=0.999$ for gHB and HB states with $N=2,4,6$. Considering $r^C_H$ for $y=1$ (phase estimation), due to the manifestation of the optimality of the double homodyne measurement, we get $r^C_H=1$. However, the value of $r^C_H$ remains below 1.5 until $y=0.5$ which indicates that, in the multiparameter setting, even though the optimality for a joint estimation is non-achievable, the measurement compatibility performs very well. On the other hand, we find that the value of $r_{BG}\approx1$ for all values of $y$.

Therefore, since the double homodyne measurement is optimal for phase estimation, one finds higher compatibility when weights are chosen closer to phase estimation. As weights are chosen closer to loss estimation, the sub-optimality or the incompatibility of the measurement in estimating loss becomes more pronounced leading to higher values of $r^C_H$ (refer to Eq.\,\ref{eqn:mceqn}). At the same time, as we are estimating extremely low values of losses, the joint-estimation of phase and loss effectively becomes a single-parameter phase estimation problem for which $C_H\approx C_Q$. However, the incompatibility term in $C_H$ becomes more pronounced when one estimates higher values of losses.
Thus, in the case of very low losses, the interplay of $C_C$, $C_H$, and $C_Q$, combined with the parameter weights, gives rise to a region where the HCRB is closely attainable with the double homodyne measurement on individual copies of the probe.

\clearpage


\bibliography{refs}

\end{document}